\documentclass[journal]{IEEEtran}
\ifCLASSINFOpdf
\else
\fi

\usepackage{amsmath, amssymb, amsfonts, amsthm, booktabs, cancel, comment, enumitem, eurosym, float, graphicx, mathtools, multirow, nicefrac, placeins, soul, steinmetz, stfloats, siunitx, svg, url, xcolor, tabularx}%
\usepackage{commath}
\usepackage{cite}
\usepackage{subfigure}
\usepackage{hyperref}
\usepackage{nomencl}
\usepackage{cleveref}
\makenomenclature

\usepackage{makecell}

\floatstyle{ruled}
\newfloat{model}{thp}{lop}
\floatname{model}{\textsc{Model}}

\makeatletter
\renewcommand*{\fnum@model}{\fname@model}
\makeatother

\begin{document}

\title{Exposing Barriers to Flexibility Aggregation\\in Unbalanced Distribution Networks} 





%
%
%

\author{
        Andrey~Churkin\raisebox{0.8ex}{\scalebox{0.7}{$\blacklozenge$}},~\IEEEmembership{Member, IEEE},
        Wangwei~Kong\raisebox{0.8ex}{\scalebox{0.7}{$\blacklozenge$}},~\IEEEmembership{Member, IEEE},\\
        
        Pierluigi~Mancarella,~\IEEEmembership{Fellow, IEEE},
        and~Eduardo~A.~Martínez~Ceseña,~\IEEEmembership{Member,~IEEE}
\thanks{\raisebox{0.2ex}{\scalebox{0.85}{$\blacklozenge$}}\,Equal contributions by Wangwei Kong and Andrey Churkin.}
\thanks{This work was carried out as a part of the ATTEST project (the Horizon 2020 research and innovation programme, grant agreement No 864298).}
\thanks{Andrey Churkin is with the Dyson School of Design Engineering, Imperial College London, U.K.}
\thanks{Wangwei Kong is with National Grid Electricity Transmission, U.K.}
\thanks{Pierluigi Mancarella is with the Department of Electrical and Electronic Engineering, the University of Manchester, U.K., and also with the Department of Electrical and Electronic Engineering, the University of Melbourne, Australia.}
\thanks{Eduardo Alejandro Martínez Ceseña is with the Department of Electrical and Electronic Engineering, the University of Manchester, U.K., and also with the Tyndall Centre for Climate Change Research, U.K.}
}

\maketitle

\begin{abstract}
%
The increasing integration of distributed energy resources (DER) offers new opportunities for distribution system operators (DSO) to improve network operation through flexibility services. To utilise flexible resources, various DER flexibility aggregation methods have been proposed, such as the concept of aggregated P-Q flexibility areas.
Yet, many existing studies assume perfect coordination among DER and rely on single-phase power flow analysis, thus overlooking barriers to flexibility aggregation in real unbalanced systems.
To quantify the impact of these barriers, this paper proposes a three-phase optimal power flow (OPF) framework for P-Q flexibility assessment, implemented as an open-source Julia tool \texttt{3FlexAnalyser.jl}. 
The framework explicitly accounts for voltage unbalance and imperfect coordination among DER in low voltage (LV) distribution networks.
Simulations on an illustrative 5-bus system and a real 221-bus LV network in the UK reveal that over 30\% of the theoretical aggregated flexibility potential can be lost due to phase unbalance and lack of coordination across phases.
These findings highlight the need for improved flexibility aggregation tools applicable to real unbalanced distribution networks.

\end{abstract}


%
\IEEEpeerreviewmaketitle

\section{Introduction}
\IEEEPARstart{M}{odern} distribution systems are evolving rapidly with the increasing integration of distributed energy resources (DER), particularly in low voltage (LV) networks \cite{Eid2016,Chen2020,Liu_Ochoa2022}. Controllable DER, such as battery energy storage systems (BESS), prosumers, and electric vehicles (EV), have the technical ability to adjust their power exchange with the grid and thus provide flexibility services, for example, by helping to balance electricity supply and demand \cite{Schittekatte2020,Riaz2022}. Such services can improve the operation of distribution systems and can also be aggregated and traded between distribution and transmission system operators (DSO and TSO) \cite{Gerard2018,Vicente-Pastor2019,Muller2019,Givisiez2020,Yi2021,Paredes2023}. 

To enable flexibility services, flexibility aggregation models are essential both at the local level (within LV feeders) and system level (at primary substations or TSO-DSO interfaces) \cite{Heleno2015,Silva2018,Capitanescu2018,Contreras2018,Kalantar-Neyestanaki2020,Tan2020,Bolfek2021,Contreras2021,Lopez2021,Petrou2021,Sarstedt2022,Fruh2022,Nazir2022,Riaz2022,Liu_Ochoa2022,ChurkinSegmentation,Churkin2024,Rabiee2024,Chrysostomou2025,liu2025region,Pozo2025,Jiang2025}.
These models aim to estimate the limits of flexible active (P) and reactive (Q) power exchanges at specific locations of a distribution network.
By estimating the P-Q limits, DSO can clear flexibility markets while managing network constraints.
The sets of all feasible power exchanges (operating points) in the P-Q space are known as aggregated P-Q flexibility areas or nodal operating envelopes \cite{Riaz2022,Churkin2024}.



Despite recent advances in DER flexibility aggregation models, several critical assumptions regarding DER operation in unbalanced LV networks remain largely unchallenged, making it easy to overlook barriers to flexibility aggregation in real-world distribution networks. Specifically, this paper exposes two barriers that can significantly reduce the usable flexibility from DER:
\begin{itemize}
\item \textbf{Barrier \#1}: Lack of DER coordination. Most flexibility aggregation models assume that all flexibility providers (referred to as flexible units in this work) can be perfectly coordinated \cite{Yi2021,Heleno2015,Silva2018,Capitanescu2018,Contreras2018,Kalantar-Neyestanaki2020,Tan2020,Bolfek2021,Contreras2021,Lopez2021,Sarstedt2022,Fruh2022,Nazir2022,Riaz2022,ChurkinSegmentation,Rabiee2024,Chrysostomou2025,liu2025region,Jiang2025,Pozo2025}. 
Such centralised aggregation methods imply that units at different locations can jointly adjust their power in order to manage voltage constraints and keep the network's operation feasible. However, in real distribution systems, flexible units connected to different phases across the network may not be perfectly coordinated due to low observability, data exchange issues, single-phase connection, behind-the-meter devices, and other constraints \cite{Churkin2024,Petrou2021}.
These practical limitations of DER coordination create a barrier to effective flexibility aggregation.
\item \textbf{Barrier \#2}: Impacts of phase unbalance.
Many existing studies rely on single-phase power flow models that implicitly assume perfectly balanced conditions in distribution networks \cite{Yi2021,Heleno2015,Silva2018,Capitanescu2018,Contreras2018,Kalantar-Neyestanaki2020,Tan2020,Bolfek2021,Contreras2021,Lopez2021,Sarstedt2022,Fruh2022,Nazir2022,Riaz2022,ChurkinSegmentation,Churkin2024,Rabiee2024,Chrysostomou2025,liu2025region,Jiang2025,Pozo2025}.
Although a few works on flexibility services do consider unbalanced distribution systems and apply three-phase optimal power flow (OPF) models, e.g. \cite{Chen2020,LEIVA2020,FOBES2020,Petrou2021,Wang2021,Avramidis2021,Liu_Ochoa2022,Russell2023,Antic2024,Liu2024,Liu2024superellipsoid,Lu2024,Zhang2025}, they do not quantify the ensuing voltage unbalances and the volume of flexibility lost due to voltage unbalance limits.
\end{itemize}

In summary, the impacts of voltage unbalance on aggregated P-Q flexibility areas remain largely unexplored, especially when combined with limited coordination among DER. Together, these factors can pose major barriers to flexibility aggregation, making a significant share of theoretical flexibility practically unusable.

These barriers are present in real distribution systems as LV networks are inherently unbalanced and the increase of single-phase uncoordinated distributed generation will further aggravate existing voltage unbalances \cite{Ma2020,mousa2024comprehensive}.
For example, suboptimal selection of DER connection phases can lead to voltage unbalance and limit DER capacity \cite{Wang2021,Antic2024}.
The need for simulating DER flexibility using accurate three-phase power flow models is acknowledged by the literature \cite{LEIVA2020,Chen2020,Avramidis2021,Liu_Ochoa2022}. However, aside from the recent work \cite{Bruno2021}, there have been no attempts to analyse the aggregated P-Q flexibility for unbalanced LV distribution networks. Note that although P-Q flexibility areas were calculated using a three-phase OPF model in \cite{Bruno2021}, the impact of voltage unbalance constraints was not explicitly quantified.
Moreover, no studies have analysed the loss of coordination between DER connected to different phases and its impacts on voltage unbalance. {\color{black}A detailed review of the existing literature and associated research gaps is provided in Section~\ref{Section: literature}.}









In this regard, this paper investigates the provision of flexibility services in unbalanced LV distribution networks and proposes a framework for quantifying the impacts of phase unbalance and imperfect coordination between distributed flexibility providers.
At the core of the framework lies an accurate nonlinear three-phase AC OPF model designed for simulating the operation of flexible units connected to different phases.
To characterise feasible flexibility services available in distribution networks, the concept of aggregated P-Q flexibility areas is applied.
Then, \textit{voltage unbalance constraints} are explicitly included to limit the deviations in phase voltages caused by flexibility services.
Finally, \textit{phase coordination constraints} are imposed to simulate situations where flexible units connected to different phases are unable to adjust their power outputs in a coordinated manner.
In mathematical optimisation terms, these constraints fix the power injections of DER in other phases, effectively removing them as control variables in the optimisation problem. 
Such constraints simulate the practical challenge of coordinating single-phase DERs connected across unbalanced phases without full observability or control.
By combining these modelling elements, the framework directly translates voltage unbalance and phase coordination constraints into reductions in the aggregated P-Q flexibility.

The models supporting the framework have been developed in Julia programming language and published as an open-source tool. The tool, named \texttt{3FlexAnalyser.jl}, and the case studies are available at \cite{3FlexAnalyser}.


To the best of the authors' knowledge, this work presents the first detailed analysis of the impacts of phase unbalance and DER coordination on aggregated P-Q flexibility in LV distribution networks.
Specifically, the paper makes the following contributions:
\begin{itemize}
\item \textbf{Contribution \#1}: A novel framework is proposed for quantifying the impacts of phase unbalance and DER coordination on aggregated flexibility in LV distribution networks. By combining three-phase AC power flow models and the concept of P-Q flexibility areas, the framework enables direct translation of voltage unbalance and phase coordination constraints into reductions in the aggregated P-Q flexibility.
\item \textbf{Contribution \#2}: New quantification analysis is performed to demonstrate that phase unbalance and lack of coordination between DER connected to different phases can lead to significant losses in aggregated DER flexibility. 
Simulations performed for an illustrative 5-bus system and a real 221-bus LV network in the UK show that over 30\% of the theoretical aggregated flexibility potential can be practically unusable.
\end{itemize}

Moreover, extensive simulations for case studies with different combinations of voltage unbalance and phase coordination constraints have yielded the following key findings:
\begin{itemize}
\item \textbf{Finding \#1}: Flexibility services in LV distribution networks can be significantly constrained due to inherent load unbalances, even in cases where flexible resources provide balanced output.
\item \textbf{Finding \#2}: Several small single-phase flexible units have the potential to provide more flexible power than a few larger three-phase balanced flexible units. However, perfect coordination of single-phase units is required to achieve this potential. 
\item \textbf{Finding \#3}: Without coordination, single-phase flexible units can exacerbate existing voltage unbalances. A stochastic analysis shows that introducing uncoordinated flexible resources frequently leads to voltage unbalance violations, occurring in over 15\% of simulated scenarios.
\item \textbf{Finding \#4}: The worst conditions for providing flexibility services include the lack of coordination between flexible units connected to different phases and tight voltage unbalance constraints. In such cases, flexible units cannot effectively manage voltage unbalance across different phases and locations, which results in the infeasibility of services provision.

\end{itemize}

The rest of the paper is organised as follows.
{\color{black}Section~\ref{Section: literature} provides a literature review by mapping the relevant research directions and representative studies.}
Section~\ref{Section: framework} introduces the modelling framework for simulating DER operation in LV unbalanced networks and quantifying the impacts of voltage unbalance and phase coordination constraints. 
In Section~\ref{Section: results}, extensive simulations are performed for an illustrative 5-bus distribution system and a real 221-bus LV network in the UK. Section~\ref{Section: discussion} further discusses the modelling aspects of flexibility aggregation. Section~\ref{Section: conclusion} concludes the paper.


\section{Literature Review}\label{Section: literature}
\begin{figure}
    \centering
    \includegraphics[width=0.94\columnwidth]{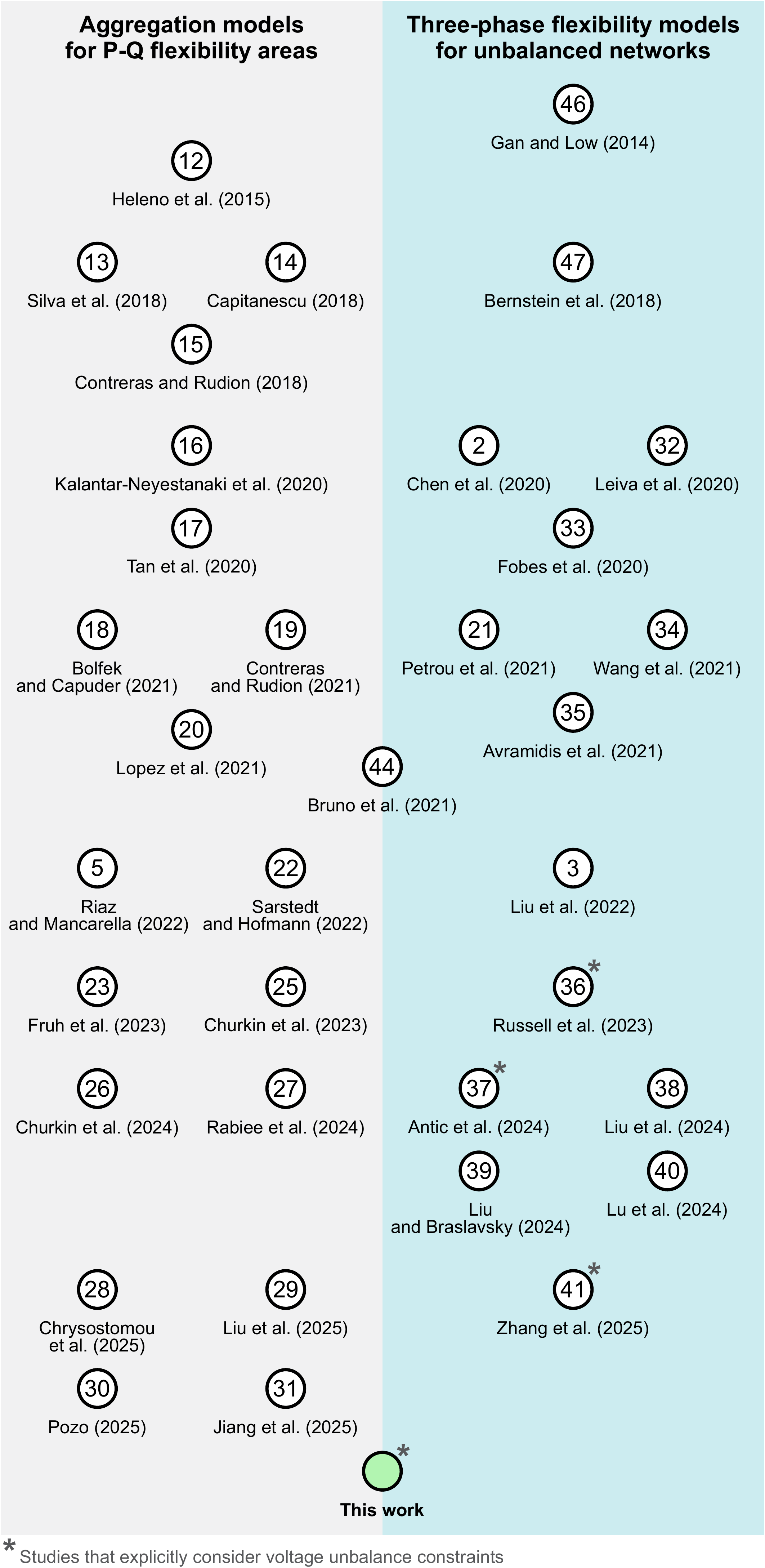}
    \caption{\color{black}Literature review mapping. Columns represent the two research directions most relevant to this work. Representative studies are located according to their contributions to these directions and are arranged chronologically, with more recent publications displayed at the bottom. Each study is visualised by a circle with the corresponding reference number. 
    }
    \label{Fig: literature review}
\end{figure}

{\color{black}
The relevant research on flexibility in distribution networks can be broadly divided into two main directions. The first research stream focuses on aggregation models based on the concept of P-Q flexibility areas \cite{Heleno2015,Silva2018,Capitanescu2018,Contreras2018,Kalantar-Neyestanaki2020,Tan2020,Bolfek2021,Contreras2021,Lopez2021,Sarstedt2022,Fruh2022,Riaz2022,ChurkinSegmentation,Churkin2024,Rabiee2024,Chrysostomou2025,liu2025region,Jiang2025,Pozo2025}. These areas characterise the feasible operating regions of distribution networks and enable the utilisation and trading of aggregated flexibility services between DSO and TSO. Most studies in this stream investigate the economic value of aggregated flexibility, exploring flexibility markets and coordination mechanisms. They rely on single-phase power flow models and therefore do not explicitly address flexibility provision in unbalanced LV networks.
The second direction considers the operation of DER in unbalanced distribution networks, relying on three-phase OPF formulations to capture phase coupling and network constraints. However, these studies mainly focus on asset-level modelling, such as the analysis of DER operating envelopes or hosting capacity, and do not estimate aggregated P-Q flexibility areas at the network level.
As a result, the integration of P-Q flexibility areas with three-phase modelling of unbalanced networks remains largely unexplored.
This work is positioned at the intersection of the two research directions and addresses this gap.
A detailed literature review mapping is shown in Fig.~\ref{Fig: literature review}.

In the research direction on aggregation models for flexibility assessment, significant efforts have been devoted to the development of algorithms for computing the boundaries of flexibility areas in the P-Q space \cite{Heleno2015,Silva2018,Capitanescu2018,Contreras2018,Bolfek2021,Contreras2021,Lopez2021}. A range of algorithms has been proposed, including random sampling and iterative optimisation-based methods.
Several studies have extended P-Q flexibility aggregation to probabilistic settings by accounting for forecast uncertainty, stochastic DER dispatch, and the reliability of flexible units \cite{Kalantar-Neyestanaki2020,Tan2020,ChurkinSegmentation}.
Another group of works focuses on the economics of flexibility provision, developing models for cost-optimal aggregation and valuation of flexibility \cite{Riaz2022,Sarstedt2022,Churkin2024,liu2025region}.
Recent studies have further explored the properties of flexibility aggregation and disaggregation, proposing data-driven models and approximation approaches \cite{Fruh2022,Rabiee2024,Chrysostomou2025,Jiang2025,Pozo2025}.
However, research in this direction relies on single-phase power flow models and does not explicitly consider flexibility provision in unbalanced LV networks.

In parallel, multiple studies have investigated the flexible operation of DER in unbalanced distribution networks. This progress has been facilitated by advances in three-phase power flow modelling \cite{Gan2014,Bernstein2018} and the release of open-source distribution power flow frameworks \cite{FOBES2020}.
Leveraging accurate three-phase power flow models, existing studies investigate the operation of DER in unbalanced distribution networks, primarily focusing on estimating network hosting capacity and DER operating limits \cite{Chen2020,LEIVA2020,Petrou2021,Wang2021,Avramidis2021,Liu_Ochoa2022,Liu2024,Liu2024superellipsoid,Lu2024}.
Several studies have examined the impact of phase unbalance on DER operation by explicitly accounting for voltage unbalance constraints \cite{Russell2023,Antic2024,Zhang2025}.
However, these works do not adopt the concept of aggregated P-Q flexibility areas and do not quantify the impact of phase coordination and voltage unbalance constraints on aggregated flexibility.

It follows that the intersection of these two research directions remains largely unexplored, as existing studies typically focus either on simplified single-phase modelling for P-Q flexibility aggregation or on detailed three-phase formulations without adopting the concept of aggregated P-Q flexibility areas.
A notable emerging exception is \cite{Bruno2021}, which demonstrates that aggregated P-Q flexibility of an unbalanced network can be computed by accounting for flexible resources and network constraints in each phase.
Yet, the impact of phase coordination and voltage unbalance constraints on three-phase P-Q flexibility areas is not explicitly quantified.
This work further advances the modelling of aggregated P-Q flexibility areas and bridges this gap.































 





}





\section{Modelling Framework}\label{Section: framework}
This section introduces the model formulation and modelling assumptions used to simulate flexibility services in three-phase unbalanced LV distribution networks. The framework consists of three main components: 1) first, an accurate nonlinear AC optimal power flow (OPF) model is selected to represent distribution networks with DER and flexible resources, 2) then, the concept of P-Q flexibility areas is adapted to quantify flexibility services available in the networks, 3) finally, voltage unbalance and phase coordination constraints are imposed to analyse their impacts on flexibility services.

To accurately simulate LV distribution networks with flexible units connected to different phases, a multi-phase flexibility estimation model \eqref{Model: objective}-\eqref{Model: phase_coordination} is formulated as a nonlinear programming problem (NLP). The formulation is based on the \textit{bus injection model} widely applied in unbalanced distribution network optimisation studies \cite{Gan2014,FOBES2020,Bernstein2018}.
Variables of the model are defined for each phase $\phi$ and include complex voltages and complex power injections at each bus $i$, $V_{i,\phi}$, $s_{i,\phi}$, complex power flows for each line $(i,j)$, $s_{ij,\phi}$, active and reactive power of generators $P_{g,i,\phi}$, $Q_{g,i,\phi}$, and active and reactive power of flexible units $P_{f,i,\phi}$, $Q_{f,i,\phi}$. 
Note that the outputs of flexible units for each phase are the primary control variables of the optimisation problem used to construct the aggregated P-Q flexibility of the network.
Voltage unbalances for the selected set of buses $\mathcal{N}^{\text{vu}}$ are explicitly modelled using the positive and negative voltage sequences, ${V_{1}}_i, {V_{2}}_i$, and voltage unbalance factors (VUF), $VUF_i$.
The original mathematical formulation of the three-phase power flow problem is adapted from the \texttt{PowerModelsDistribution.jl} package \cite{FOBES2020}. 
{\color{black}This three-phase AC OPF formulation inherently captures the propagation of unbalanced voltages and currents along network lines and allows the explicit modelling of single-phase, two-phase, and three-phase load connections.}\footnote{\color{black}In this work, network data is prepared using the OpenDSS format with a three-phase three-wire AC OPF formulation. The neutral conductor and grounding are not explicitly modelled. Investigating four-wire network modelling is left for future research.}
It is then extended by including flexible units, voltage unbalance limits, and phase coordination constraints.

\begin{figure*}
    \centering
\includegraphics[width=0.95\linewidth]{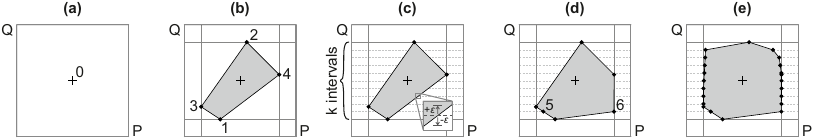}
    \caption{{\color{black}Conceptual} step-by-step illustration of the interval-based OPF algorithm for constructing an aggregated P-Q flexibility area: (a) the initial operating point with no flexibility is calculated; (b) the bounds of the area are obtained by calculating Q minimum and maximum, points 1 and 2, and P minimum and maximum, points 3 and 4; (c) these limits are discretised into intervals – 10 Q intervals in this example; (d) the first Q interval is considered by constraining the Q component of the aggregated flexibility, the corresponding P limits (boundary points 5 and 6) are calculated; (e) after evaluating all intervals, the final area is estimated as the hull of all boundary points.
    The algorithm is executed separately for each phase.
    {\color{black}The P and Q axes in this figure are schematic and do not correspond to specific numerical values. In real calculations for distribution networks, these axes can be expressed in kW and kVAr or in per-unit values.}
    }
    \label{Fig: iterative PQ algorithm}
\end{figure*}

The objective function \eqref{Model: objective} minimises or maximises the network’s active and reactive power injections at the reference bus, where coefficients $\alpha^p_i$ and $\alpha^q_i$ are introduced to control the optimisation directions in the P-Q space. In this formulation, reference bus $i^{\text{ref}}$ is the location for which flexibility from multiple flexible units is aggregated. Thus, the objective function with adjustable coefficients $\alpha^p_i$, $\alpha^q_i$ allows to iteratively (point by point) estimate the limits of aggregated flexibility in the P-Q space at the reference bus \cite{Churkin2024,ChurkinSegmentation}.
The role of the objective function \eqref{Model: objective} in algorithms for constructing P-Q flexibility areas is further discussed later in this section, with a step-by-step illustration provided in Fig.~\ref{Fig: iterative PQ algorithm}.
Equations \eqref{Model: V_min_max}-\eqref{Model: qf_min_max} represent the nonlinear AC OPF constraints. Specifically, voltage limits are imposed in \eqref{Model: V_min_max}, complex power flows and bus injections are defined in \eqref{Model: s_ij}, \eqref{Model: s_i}, where $Y_{ij}$ is the admittance matrix. 
Active and reactive power balance constraints for each node are introduced in \eqref{Model: balance_p}, \eqref{Model: balance_q}, where $d \in \mathcal{D}$ represents loads of the network. Line capacity limits are included in \eqref{Model: s_ij_max}. Finally, active and reactive power of generators and flexible units is limited in \eqref{Model: pg_min_max}-\eqref{Model: qf_min_max}.

\begin{model}[t]
\caption{{{\color{black}Steady-state} multi-phase P-Q flexibility estimation}} 
\label{Model2}
\begin{subequations} 
\label{Mod: TEP1}
\vspace{-3\jot}
\begin{IEEEeqnarray}{lll}
    \textbf{Variables:} \text{ (for each phase $\phi \in \{a,b,c\}$)} & \IEEEnonumber\\
    V_{i,\phi}  & i \in \mathcal{N}\IEEEnonumber\\
    s_{i,\phi}  & i \in \mathcal{N}\IEEEnonumber\\
    s_{ij,\phi}  & (i,j) \in \mathcal{L}\IEEEnonumber\\
    P_{g,i,\phi},Q_{g,i,\phi} & i \in \mathcal{N}, g \in \mathcal{G} \IEEEnonumber\quad\\
    P_{f,i,\phi},Q_{f,i,\phi} & i \in \mathcal{N},f \in \mathcal{F} & \IEEEnonumber \\
    {V_{1}}_i, {V_{2}}_i, {VUF}_i  & i \in \mathcal{N}^{\text{vu}} & \IEEEnonumber  \vspace{1\jot}\\
    {\textbf{Objective:}} & \IEEEnonumber\\
    \min \enskip \alpha^p_i \Re(s_{i,\phi}) + \alpha^q_i \Im(s_{i,\phi}) \quad & i = i^{\text{ref}}\quad \label{Model: objective} \vspace{1\jot}\\
    {\textbf{AC OPF constraints:}} & \IEEEnonumber\\
    V^{\min} \leq \left| V_{i,\phi} \right| \leq V^{\max} & \forall i \in \mathcal{N} \label{Model: V_min_max}\\
    s_{ij,\phi} = \left({\mathrm{diag}\left\lbrack V_{i}\left( V_{i} - V_{j} \right)^{H}{Y_{ij}}^{H} \right\rbrack} \right)_\phi \quad & \forall (i,j) \in \mathcal{L} \quad \label{Model: s_ij}\\
    s_{i,\phi} = \sum_{(i,j) \in \mathcal{L}} s_{ij,\phi} & \forall i \in \mathcal{N} \quad \label{Model: s_i}\\
    \sum_{g ,d ,f}^{\mathcal{G},\mathcal{D},\mathcal{F}}(P_{g,i,\phi} - P_{d,i,\phi} + P_{f,i,\phi})= \Re(s_{i,\phi}) & \forall i \in \mathcal{N} \label{Model: balance_p}\\
    \sum_{g ,d ,f}^{\mathcal{G},\mathcal{D},\mathcal{F}}(Q_{g,i,\phi} - Q_{d,i,\phi} + Q_{f,i,\phi})= \Im(s_{i,\phi}) \enskip\enskip & \forall i \in \mathcal{N} \label{Model: balance_q}\\
    \left|s_{ij,\phi}\right| \leq s_{ij,\phi}^{\max} & \forall (i,j) \in \mathcal{L} \label{Model: s_ij_max}\\
    P_{g,i,\phi}^{\min} \leq P_{g,i,\phi} \leq P_{g,i,\phi}^{\max} & \forall g \in \mathcal{G} \label{Model: pg_min_max}\\
    Q_{g,i,\phi}^{\min} \leq Q_{g,i,\phi} \leq Q_{g,i,\phi}^{\max} & \forall g \in \mathcal{G} \label{Model: qg_min_max}\\
    P_{f,i,\phi}^{\min} \leq P_{f,i,\phi} \leq P_{f,i,\phi}^{\max} & \forall f \in \mathcal{F} \label{Model: pf_min_max}\\
    Q_{f,i,\phi}^{\min} \leq Q_{f,i,\phi} \leq Q_{f,i,\phi}^{\max} & \forall f \in \mathcal{F} \label{Model: qf_min_max} \vspace{1\jot}\\
    {\textbf{Voltage unbalance limits:}} & \IEEEnonumber\\
    {V_{1}}_i = \frac{{V_{i,\{a\}}} + \boldsymbol{a}{V_{i,\{b\}}} + \boldsymbol{a}^2{V_{i,\{c\}}}}{3} & i \in \mathcal{N}^{\text{vu}} \label{Model: v1}\\
    {V_{2}}_i = \frac{{V_{i,\{a\}}} + \boldsymbol{a}^2{V_{i,\{b\}}} + \boldsymbol{a}{V_{i,\{c\}}}}{3} & i \in \mathcal{N}^{\text{vu}} \label{Model: v2}\\
    VUF_i = \frac{{V_{2}}_i}{{V_{1}}_i} \times 100\% & i \in \mathcal{N}^{\text{vu}} \label{Model: vuf}\\
    VUF_i \leq \overline{VUF} & i \in \mathcal{N}^{\text{vu}} \label{Model: vuf_lim} \vspace{1\jot}\\
    {\textbf{Phase coordination constraints:}} & \IEEEnonumber\\
    \sum_{i \in \mathcal{N},f \in \mathcal{F}}P_{f,i,\varphi} = 0, \enskip \sum_{i \in \mathcal{N},f \in \mathcal{F}}Q_{f,i,\varphi} = 0 & \forall \varphi \neq \phi \label{Model: phase_coordination}
    \vspace{-1\jot}
\end{IEEEeqnarray}
\end{subequations}
\end{model}

The formulated AC OPF model \eqref{Model: objective}-\eqref{Model: qf_min_max} enables estimating the limits of flexibility services for the selected phase and bus of a distribution network, subject to available flexible resources and network constraints.
However, this model assumes perfect coordination of the flexible resources connected to different phases. Under this assumption, all flexible units can act together to maximise a flexibility service at the reference bus while managing network constraints. Moreover, the model does not explicitly constrain voltage unbalance caused by the operation of flexible units.
To investigate the impacts of phase unbalance on the provision of flexibility services, voltage unbalance constraints \eqref{Model: v1}-\eqref{Model: vuf_lim} are explicitly included in the formulation.
In \eqref{Model: v1}, \eqref{Model: v2}, the positive sequence voltage and the negative sequence voltage are defined for the selected set of buses $\mathcal{N}^{\text{vu}}$, where $\boldsymbol{a} = 1 \phase{120^{\circ}}$ \cite{PERERA2023,nakadomari2024}. 
In \eqref{Model: vuf}, VUF is defined as the ratio between the negative and positive sequences,
{\color{black}while a discussion of other voltage unbalance definitions is provided in Section~\ref{Section: discussion}.}
Then, the voltage unbalance factor is constrained in \eqref{Model: vuf_lim} by a threshold parameter $\overline{VUF}$.
The selection of this parameter depends on regional regulations and grid codes.
For example, in the UK, the regulatory VUF threshold for steady-state operation is 1.3\% \cite{EnergyNetworksAssociation1990,NationalGrid2014}.
Thus, the inclusion of constraints \eqref{Model: v1}-\eqref{Model: vuf_lim} makes it possible to limit the potential voltage unbalance caused by flexibility services. But, as will be shown by simulations, limiting voltage unbalance causes a reduction in the flexibility service capacity.


To investigate the impact of coordination between flexible resources connected to different phases, phase coordination constraints \eqref{Model: phase_coordination} are introduced. 
These constraints fix the power injections of DER connected to other phases, effectively removing them as control variables in the optimisation problem.
Specifically, in \eqref{Model: phase_coordination}, flexible units not connected to the selected phase $\phi$, that is, $\forall \varphi \neq \phi$, are not allowed to provide flexibility services and support the operation of units connected to phase $\phi$.
The inclusion of these constraints models the lack of coordination between flexible units connected to different phases, which can exacerbate voltage unbalance in the network.

To characterise the flexibility services provided by LV distribution networks, the concept of P-Q flexibility areas is applied to the context of multi-phase flexibility provision. 
Mathematically, the operation limits of multiple flexible units can be described using the Minkowski addition of their P-Q capability sets \cite{Riaz2022,AGARWAL2002,Churkin2024}.
That is, the limits of the aggregated DER flexibility in the P-Q space provided at the reference bus can be estimated as the Minkowski addition of individual flexibility limits of all available flexible units.
{\color{black}However, Minkowski addition aggregates the individual P-Q capability sets purely geometrically and cannot incorporate physical network constraints, such as line capacity and voltage limits. Moreover, the computational cost of exact Minkowski addition is known to have poor scalability due to its geometric complexity. Calculations for convex sets grow with the number of sets being summed, and addition of non-convex sets is computationally intractable in general.
}

Therefore, to accurately estimate the aggregated DER flexibility considering network constraints, the formulated optimisation model \eqref{Model: objective}-\eqref{Model: phase_coordination} can be solved iteratively, and the boundary of the network flexibility area at the reference bus can be estimated with the desired level of granularity.
Multiple algorithms have been proposed for constructing the boundaries of flexibility areas, including Monte Carlo simulations and various optimisation methods \cite{Lopez2021,Bolfek2021,Contreras2021}. 
In this work, the $\epsilon$-constraint method is chosen as a simple and straightforward algorithm for approximating the P-Q flexibility area by considering $k$ intervals \cite{Capitanescu2018,Churkin2024,ChurkinSegmentation}. 
At each interval, the feasible P-Q space of the aggregated DER flexibility at the reference bus $i^{\text{ref}}$ is constrained by inequalities \eqref{k_interval_P} or \eqref{k_interval_Q}.
\begin{subequations} 
\begin{IEEEeqnarray}{ll}
    p_{i,\phi}^k - \epsilon \leq \Re(s_{i,\phi}) \leq p_{i,\phi}^k + \epsilon \qquad & i = i^{\text{ref}}, \phi = \phi^{\text{ref}}
    \quad \label{k_interval_P}\\
    q_{i,\phi}^k - \epsilon \leq \Im(s_{i,\phi}) \leq  q_{i,\phi}^k + \epsilon \qquad & i = i^{\text{ref}}, \phi = \phi^{\text{ref}} \quad \label{k_interval_Q}
\end{IEEEeqnarray}
\end{subequations}

These inequalities fix P or Q components of the aggregated flexibility (with a negligibly small margin $\epsilon$) to specific values of interval $k$, $p_{i,\phi}^k$ or $q_{i,\phi}^k$. Then, model \eqref{Model: objective}-\eqref{Model: phase_coordination} can be solved twice to find the maximum and minimum of the non-fixed flexibility component for this interval.
By iteratively imposing the $\epsilon$ constraints and solving model \eqref{Model: objective}-\eqref{Model: phase_coordination}, the boundary of the P-Q flexibility area can be calculated as a piece-wise linear approximation with $2k$ operating points.
Fig.~\ref{Fig: iterative PQ algorithm} provides a visual overview of the interval-based iterative OPF algorithm for constructing the boundary of P-Q flexibility area.
{\color{black}
In contrast to Minkowski addition, this method incorporates network constraints and offers direct control over the accuracy (granularity) of the aggregated flexibility estimation.}
With larger $k$, the accuracy of P-Q flexibility area estimation increases, but the computational cost also grows as higher numbers of AC OPF problems have to be solved.
{\color{black}The computational complexity and scalability of this method are further discussed in Section~\ref{Section: discussion}.}

\begin{figure}
    \centering
    \includegraphics[width=0.90\columnwidth]{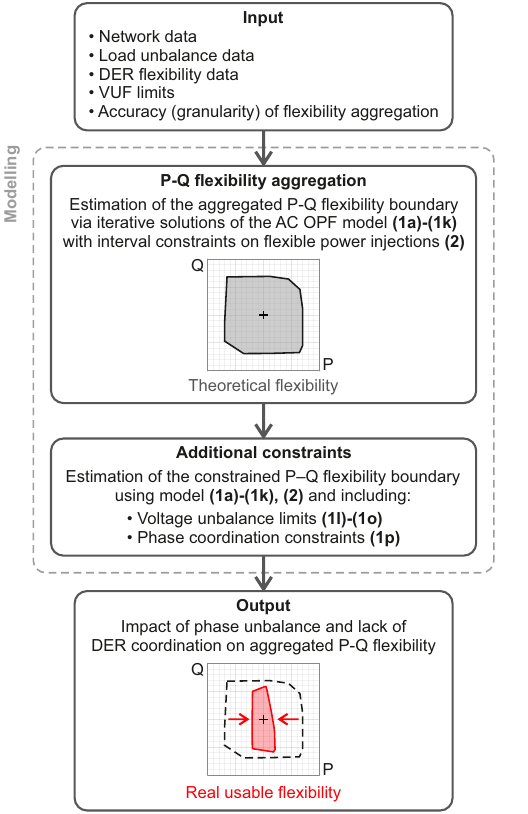}
    \caption{Flowchart of the proposed framework. Flexibility services from DER are estimated for each phase using the concept of P-Q flexibility areas. Then, the impact of phase unbalance and DER coordination is quantified by introducing corresponding constraints, resulting in reduced flexibility areas.}
    \label{Fig: new vertical flowchart}
\end{figure}

In this work, the concept of flexibility areas is extended to a multi-phase context: flexibility service limits are estimated in the P-Q space individually for each phase.
This requires selecting a reference phase $\phi^{\text{ref}}$ for which the limits of P-Q flexibility will be calculated. 
The multi-phase formulation enables quantifying the impacts of voltage unbalance and phase coordination constraints.
A high-level overview of the proposed framework is presented in Fig.~\ref{Fig: new vertical flowchart}.
The inputs include network data, available flexible resources, load and flexibility unbalances, VUF limit, and phase coordination assumptions.
Then, the limits of flexibility services from DER are estimated using the concept of P-Q flexibility areas. Next, voltage unbalance and phase coordination constraints are imposed and the P-Q flexibility areas are estimated again.
Finally, the aggregated flexibility is compared for cases with and without these constraints and the impacts of phase unbalance and DER coordination are quantified (as reductions in the aggregated P-Q flexibility areas, e.g., in kVAr\textsuperscript{2}).

{\color{black}
Note that the proposed framework is formulated as a steady-state flexibility aggregation problem at a single time instance. This formulation is intentionally selected to isolate and clearly demonstrate the impact of voltage unbalance and phase coordination constraints on aggregated flexibility. Extensions to multi-period formulations are discussed in Section~\ref{subsection: multiperiod}.
}

\section{Case Studies}\label{Section: results}
In this section, the proposed framework is demonstrated using two case studies: (i) a simple 5-bus system with DER flexibility available at a single location; (ii) a real 221-bus distribution network in the UK with 12 flexible units. For each system, the limits of flexibility services are characterised via the concept of aggregated P-Q flexibility areas. Then, the impacts of voltage unbalance and phase coordination are quantified. 
Finally, a stochastic analysis is performed to evaluate how frequently voltage unbalance violations can occur due to uncoordinated flexible resources.

{\color{black}
Data for these case studies is publicly available at \cite{churkin_Zenodo,3FlexAnalyser} in the OpenDSS network file format. These files can be directly used in the developed \texttt{3FlexAnalyser.jl} tool to reproduce the results presented in this section. The repository also provides a description of the input-output data structure, key modelling parameters, and the required Julia packages.}

{\color{black}Note that in the considered case studies, thermal loading constraints of lines and transformers are not binding for the analysed operating points and flexibility ranges. The observed reductions in aggregated flexibility are therefore driven by voltage constraints rather than by line congestion.
It is also assumed that the upstream network (supply source) is balanced at the point of connection. This assumption isolates the impact of downstream network constraints and voltage unbalance. Introducing upstream unbalance conditions would shift the feasible operating region, affecting the aggregated P-Q flexibility areas differently in each phase.


All simulations were performed using the JuMP modelling language in Julia and solved with the nonlinear optimisation solver IPOPT. This solver is widely used in the power systems literature for solving large-scale nonlinear OPF problems \cite{Alizadeh2022}. For the considered case studies, IPOPT exhibited reliable convergence across the feasible operating points. While other solvers and flexibility aggregation algorithms can be used, they are not expected to alter the qualitative insights of this study.
}

\begin{figure}
    \centering
    \includegraphics[width=0.95\columnwidth]{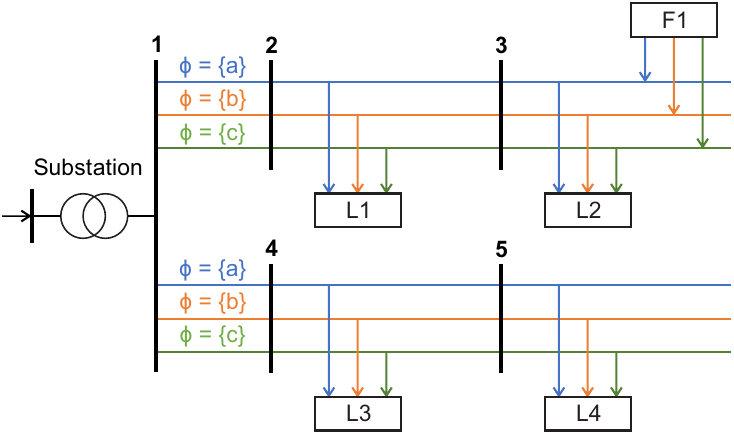}
    \caption{Case study: illustrative 5-bus system with four loads and one flexible unit. Each load can consume different power across the phases, thus creating phase unbalance. The flexible unit is assumed to produce a balanced output (to simulate a three-phase flexible device) or an unbalanced output (to simulate three independent flexibility providers connected to different phases).}
    \label{Fig: new 5-bus scheme}
\end{figure}

\subsection{Case Study: Illustrative 5-bus System}
To illustrate the principles of flexibility services estimation considering phase unbalance, a simple network with 5 buses is constructed (see Fig.~\ref{Fig: new 5-bus scheme}). The network has four loads and one flexibility provider. Each load has connections to three phases, which enables to simulate balanced network operation or to create phase unbalance due to uneven allocation of single-phase loads.
The flexible unit also has connections to all phases and therefore can be used to simulate balanced or unbalanced flexible power output.

The following network conditions are considered in simulations for this case study:
\begin{itemize}
\item \textit{Balanced network}: all loads are equal to 5 kW and are perfectly balanced, that is,  $P_{d,i,\phi} = 5.0$ $\forall i \in \mathcal{N}, \phi \in \{a,b,c\}$. The load power factor is assumed to be constant and equals 0.91.
\item \textit{Unbalanced network}: load L2 has different active power consumption per phase, \{7.0, 4.5, 3.5\} kW, with a power factor of 0.91. All other loads are equal to 5 kW and are perfectly balanced.
\item \textit{Balanced flexibility}: the flexible unit is a three-phase device producing the same flexible power output per each phase, with limits ±8 kW and ±8 kVAr, that is, $P_{f,\phi} \in [-8.0,8.0],$ $Q_{f,\phi} \in [-8.0,8.0]$ $ \forall \phi \in \{a,b,c\},$ $P_{f,\{a\}}=P_{f,\{b\}}=P_{f,\{c\}},$ $Q_{f,\{a\}}=Q_{f,\{b\}}=Q_{f,\{c\}}$.
\item \textit{Unbalanced flexibility}: three independent single-phase flexible units are connected at bus 3. These units can have different active and reactive power outputs (with limits ±8 kW and ±8 kVAr), thus providing unbalanced flexibility.
\end{itemize}

\begin{figure}
    \centering
    \includegraphics[width=0.63\columnwidth]{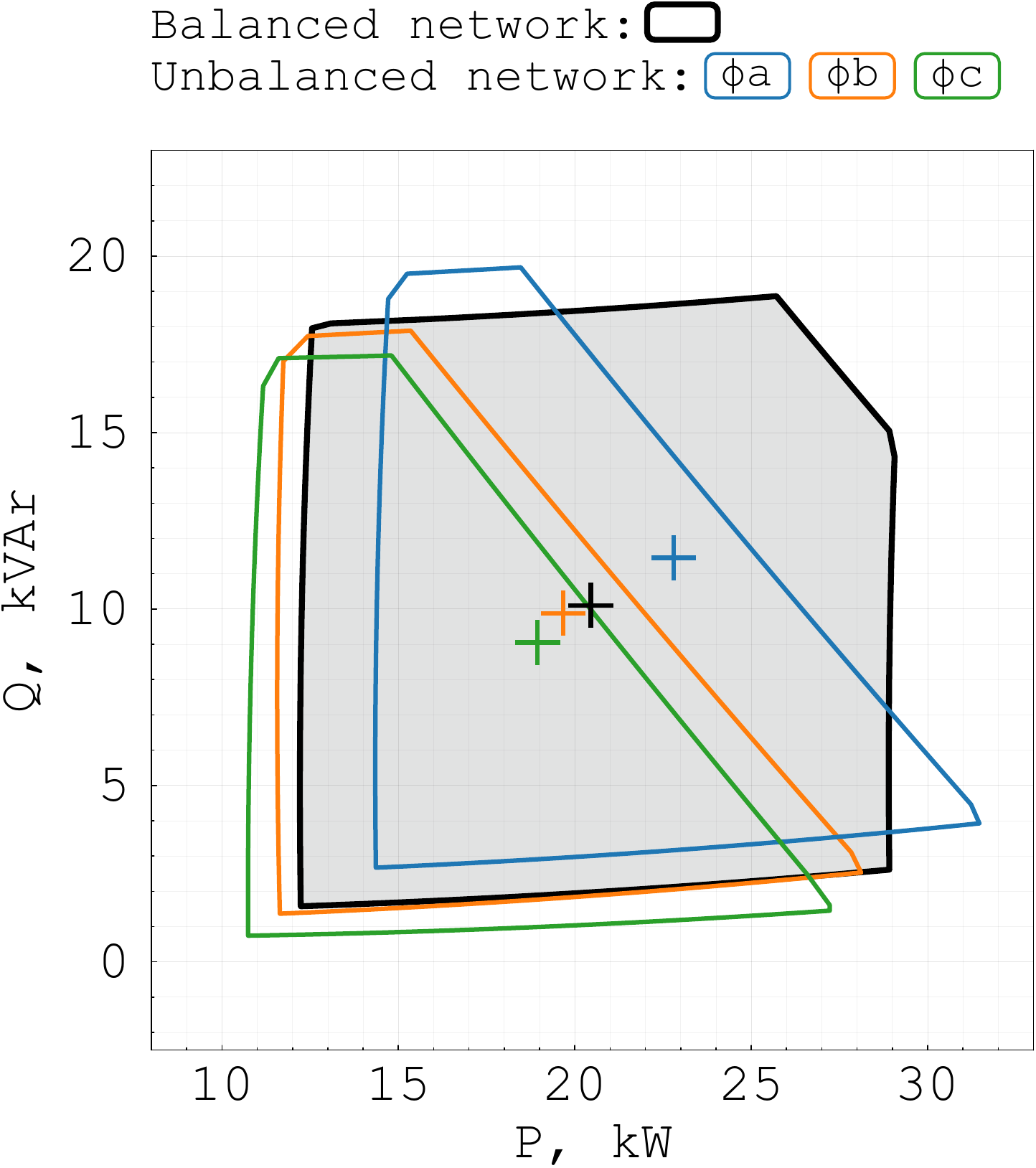}
    \caption{Analysis of the load unbalance impact on flexibility services provided by the three-phase balanced flexible unit in the 5-bus system. The services are characterised by the aggregated network flexibility area in the P-Q space at the primary substation. The cross markers correspond to the initial operating point (with no flexibility provision), while the coordinates represent the total network’s power consumption.}
    \label{fig: 5bus load unbalance analysis}
\end{figure}

First, the impact of load unbalance on flexibility services is analysed. The flexible unit is assumed to be a three-phase device producing the same flexible power output per phase. That is, by design, this unit can provide only balanced flexibility services. The question is, does the load unbalance, inherent in LV distribution networks, affect such services? To address this question, the aggregated P-Q flexibility area of the network was calculated for cases with balanced and unbalanced loads, as shown in Fig.~\ref{fig: 5bus load unbalance analysis}. Each boundary was approximated by 80 points, i.e., the flexibility estimation model \eqref{Model: objective}-\eqref{Model: phase_coordination} was iteratively solved 80 times.
In the case of a balanced network, the flexibility service is less constrained and has the largest P-Q area (displayed in grey colour).\footnote{Note that for a perfectly balanced network, there is no need to display flexibility areas per each phase as these areas coincide.}
In the case of an unbalanced network, the operation of the flexible unit is constrained differently in each phase. Therefore, the limits of the flexibility service are reduced, and the P-Q areas become different across the three phases (as highlighted in blue, orange, and green colours). The quantification of the flexibility reduction due to load unbalance is given in Table~\ref{table: load unbalance impacts 1}, where the areas of aggregated flexibility are calculated in kVAr\textsuperscript{2}. The results demonstrate that the areas are significantly reduced due to the introduced load unbalance (36-38\% reduction for the 5-bus system). It follows that balanced flexibility services provided by three-phase devices can be constrained by load unbalance present in distribution networks.

\begin{figure}
    \centering
    \includegraphics[width=0.7\columnwidth]{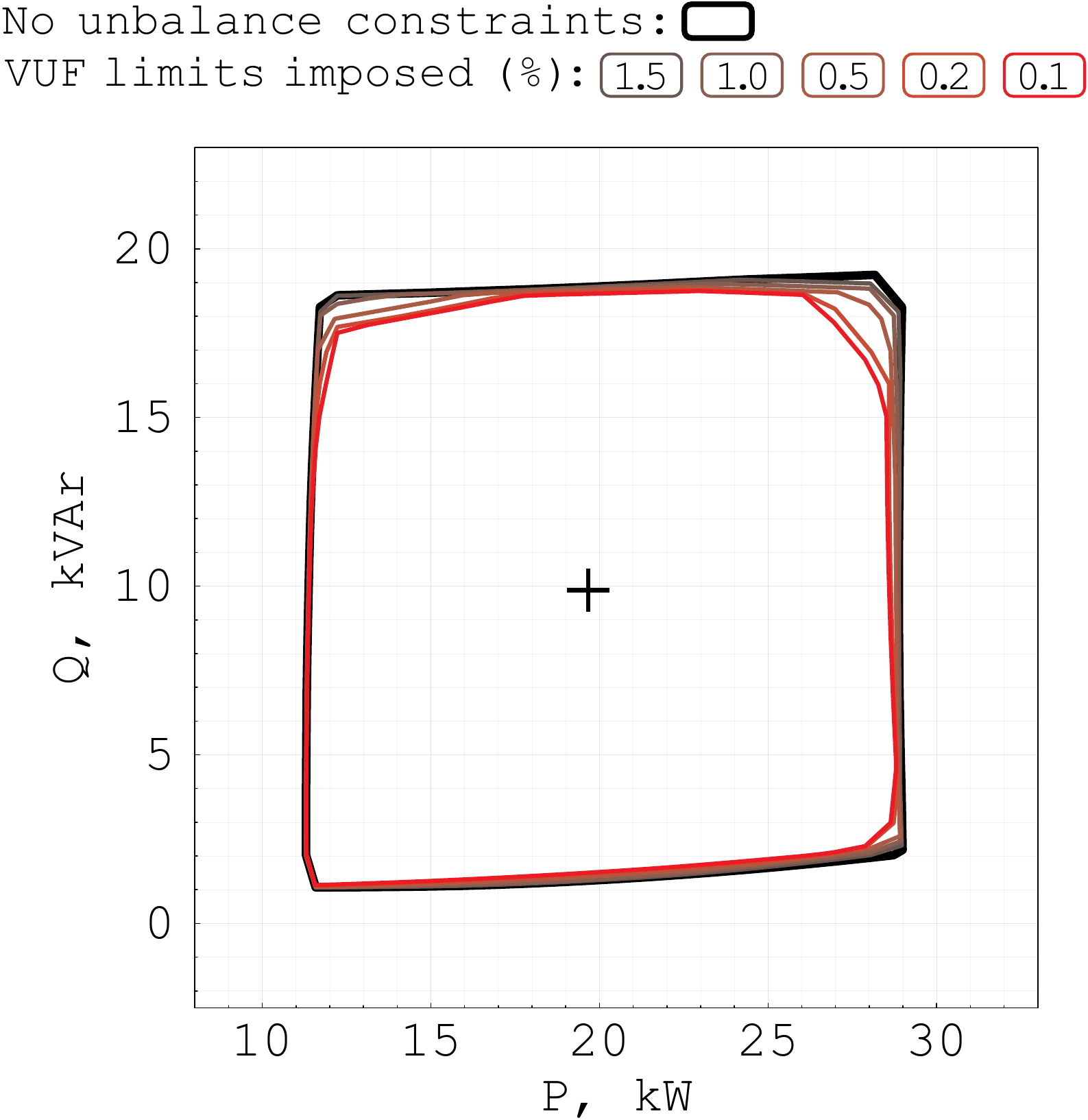}
    \caption{Analysis of the voltage unbalance impact on flexibility services in the 5-bus system, for phase $\phi= \{b\}$. The cross marker corresponds to the initial operating point (with no flexibility provision), while the coordinates represent the total network’s power consumption. The services are provided by three single-phase flexible units connected to different phases. All units are assumed to be perfectly coordinated.}
    \label{fig: 5bus VUF analysis}
\end{figure}

Second, the impact of voltage unbalance limits on flexibility services is analysed. An unbalanced network with three single-phase flexible units is considered. That is, both loads cause phase unbalance and the units can provide unbalanced flexibility services. Then, the VUF constraints \eqref{Model: v1}-\eqref{Model: vuf_lim} are imposed to control voltage unbalance caused by the uneven allocation of single-phase loads and unbalanced output of single-phase flexible units. The aggregated P-Q flexibility areas for VUF in the range 0.1-1.5\% are shown in Fig.~\ref{fig: 5bus VUF analysis}.\footnote{Note that VUF limits below 1\% are unrealistically low – no grid codes regulate such low voltage unbalances. In this work, low VUF limits are included in the analysis for the sake of illustrating the principles of flexibility provision under phase unbalance.}\textsuperscript{,}\footnote{In the 5-bus system, VUF constraints \eqref{Model: v1}-\eqref{Model: vuf_lim} are imposed on each bus.}
For simplicity, only phase $\phi=\{b\}$ is displayed in the figure – flexibility services in other phases have similar dependencies on VUF constraints.

\begin{table}
\caption{Load Unbalance Impacts on Aggregated P-Q Flexibility Provided by a Balanced Three-phase Unit in the 5-bus System}
\centering
\begin{tabular}{@{}lll@{}} 
\toprule
Simulations for balanced flexibility & \multicolumn{2}{l}{\begin{tabular}[c]{@{}l@{}} Aggregated P-Q flexibility area,\\ kVAr\textsuperscript{2} (\%)\end{tabular}}
 \\
\midrule
Balanced network, any phase & 267.96 & (100.00\,\%) \\
Unbalanced network, $\phi= \{a\}$ & 177.03 & (\phantom{1}66.07\,\%)\\
Unbalanced network, $\phi= \{b\}$ & 165.86 & (\phantom{1}61.90\,\%)\\
Unbalanced network, $\phi= \{c\}$ & 166.16 & (\phantom{1}62.01\,\%)
\\
\bottomrule
\end{tabular}
\label{table: load unbalance impacts 1}
\end{table}

Note that the area displayed in Fig.~\ref{fig: 5bus VUF analysis} is larger than the areas for balanced flexibility services analysed in Fig.~\ref{fig: 5bus load unbalance analysis} and Table~\ref{table: load unbalance impacts 1}. The three single-phase flexible units can provide more flexibility (with an area of 306.04 kVAr\textsuperscript{2}) than a single three-phase balanced unit of the same capacity. It follows that, if coordinated, multiple single-phase units can outperform three-phase units in maximising the limits of aggregated flexibility. It is also worth noting that VUF limits do not have a major impact on the flexibility provided by multiple single-phase units. Even under the extreme unrealistic VUF limit of 0.1\%, the P-Q flexibility area (highlighted in red) decreases by only 6.91\%. This happens due to the assumption of perfect unit coordination. Under this assumption, all units jointly manage voltage constraints and maximise flexibility services: any unbalance constraining flexibility in one phase can be mitigated by coordinated actions of units in other phases.

\begin{figure}
    \centering
    \includegraphics[width=0.9\columnwidth]{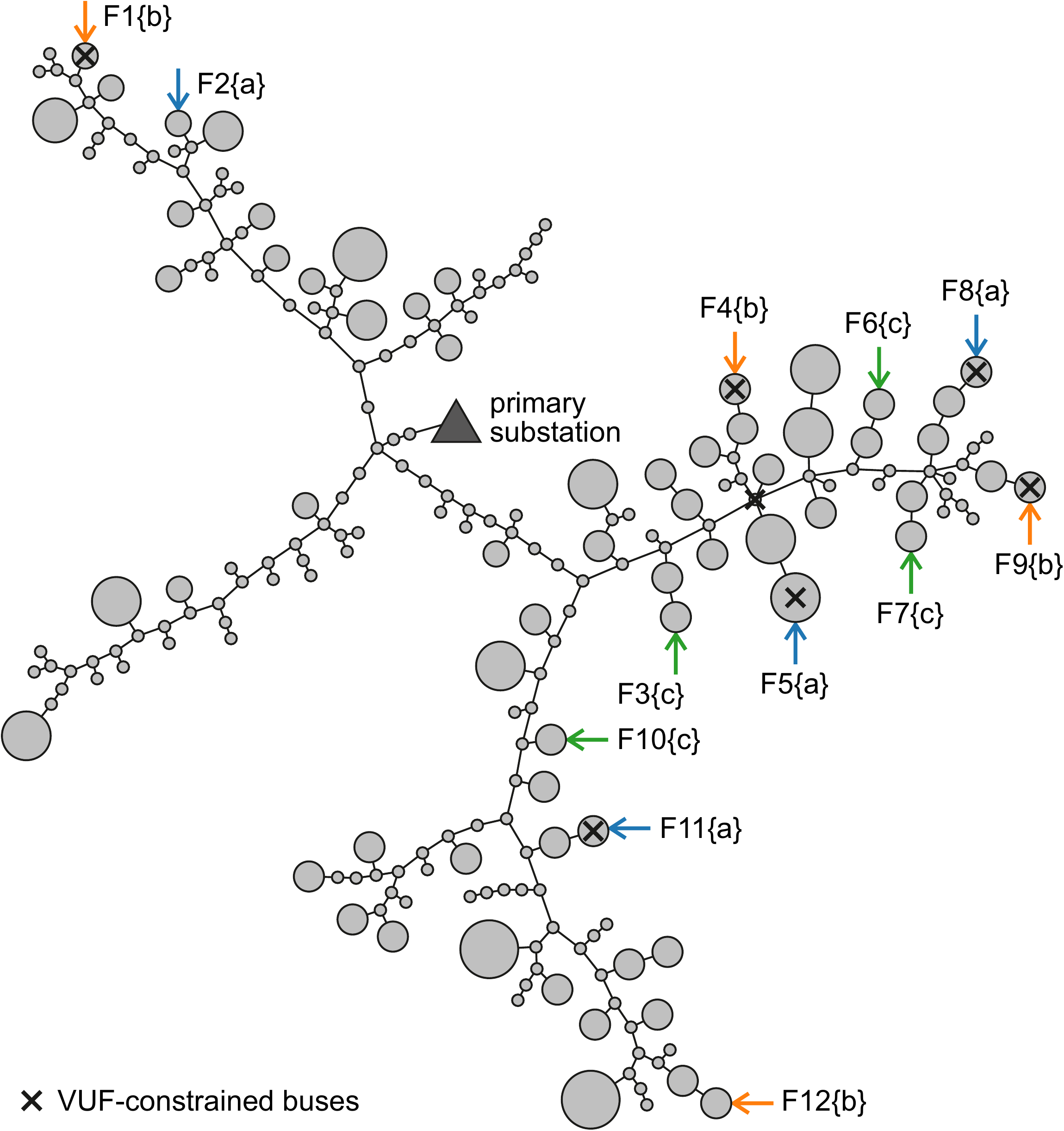}
    \caption{Case study: real 221-bus LV distribution network in the UK. The system has 12 single-phase flexible units (4 units per phase) as highlighted by the corresponding colours.}
    \label{Fig: 221bus scheme}
\end{figure}

\begin{table}
\caption{Unbalanced Load Data for the 221-bus LV Network}
\centering
\begin{tabular}{@{}lcccc@{}}
\toprule 
\ & Total & Phase $\phi= \{a\}$ & Phase $\phi= \{b\}$ & Phase $\phi= \{c\}$\\
\midrule
P, kW & 82.00 & 37.00 & 24.00 & 21.00 \\
Q, kVAr & 64.55 & 31.41 & 17.06 & 16.08 \\
\bottomrule
\end{tabular}
\label{table: UK network data}
\end{table}

\begin{figure*}
    \centering
    \includegraphics[width=0.98\textwidth]{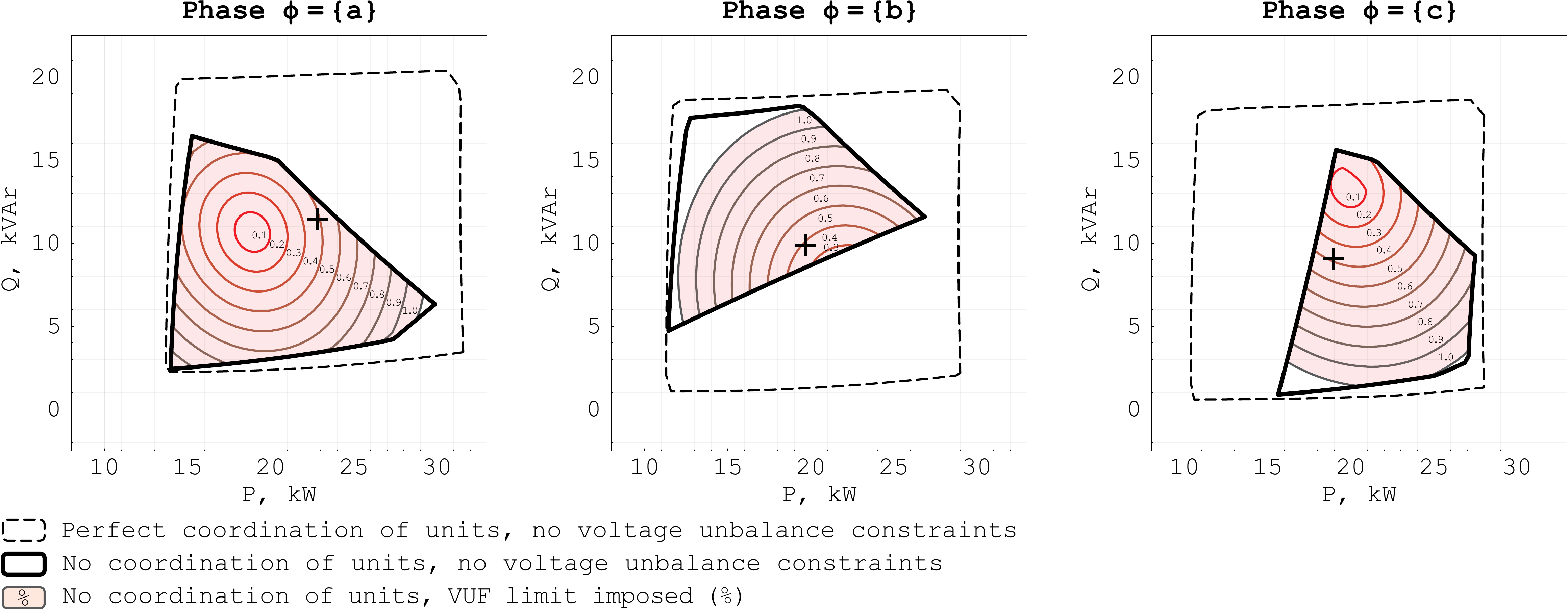}
    \caption{Impact of phase coordination and voltage unbalance constraints on flexibility services in the 5-bus system. The aggregated P-Q flexibility areas are shown for each phase. Cross markers indicate the initial operating point with no flexibility provision, while the coordinates represent the total network power consumption per phase. Flexibility services are provided by three single-phase units. Imposing phase coordination constraints prevents joint flexibility provision by units connected to different phases, thereby limiting voltage unbalance management and reducing the feasible area for each phase.
    }
    \label{fig: 5bus coordination analysis}
\end{figure*}

However, for existing distribution systems, the assumption of perfect coordination may be unrealistic because: 1) single-phase flexible units can be connected at different locations and belong to different owners, and 2) required automation and communication technologies are not widely adopted. To analyse the lack of unit coordination, phase coordination constraint \eqref{Model: phase_coordination} is imposed. Under this constraint, flexibility service in phase $\phi$ can be provided only by single-phase units connected to that phase. Other units cannot manage the voltage unbalance or assist in maximising network flexibility. The impact of phase coordination and voltage unbalance constraints on P-Q flexibility areas is shown in Fig.~\ref{fig: 5bus coordination analysis}. {\color{black}The red shaded areas correspond to simulations in which VUF constraints \eqref{Model: vuf_lim} in the range 0.1-1.0\% are iteratively imposed. 
The quantitative results for these simulations are summarised in Table~\ref{table: load unbalance impacts 2}, where aggregated flexibility areas are reported in kVAr\textsuperscript{2}. Relative reductions in flexibility are expressed in percent per phase, with 100\% denoting the maximum flexibility under perfect coordination and no voltage unbalance constraints.}
The findings of these simulations are twofold: (i) without coordination between units connected to different phases, aggregated flexibility areas reduce significantly, by over 54\%, depending on the phase; (ii) voltage unbalance limits have a major impact on flexibility services when coordination between units is absent. It follows that a lack of phase coordination combined with voltage unbalance limits creates the worst conditions for providing flexibility services: single-phase units cannot effectively manage voltage unbalance across the network, which can make flexibility provision infeasible.

\begin{table*}
\caption{\color{black}Impact of Phase Coordination and Voltage Unbalance Constraints on Aggregated P-Q Flexibility in the 5-Bus System}
\centering
\begin{tabular}{@{}l
S[table-format=3.2] @{\hspace{0.6em}(} S[table-format=3.2] @{\,\%)\hspace{1.8em}}
S[table-format=3.2] @{\hspace{0.6em}(} S[table-format=3.2] @{\,\%)\hspace{1.8em}}
S[table-format=3.2] @{\hspace{0.6em}(} S[table-format=3.2] @{\,\%)}@{}}
\toprule
 \multirow{2}{*}[-0.25em]{Simulation} & \multicolumn{6}{c}{Aggregated P-Q flexibility area per phase, kVAr\textsuperscript{2} (\%)}\\
  \cmidrule(l){2-7} 
 & \multicolumn{2}{c}{\hspace{-0.9em}Phase $\phi= \{a\}$} & \multicolumn{2}{c}{\hspace{-1.8em}Phase $\phi= \{b\}$} & \multicolumn{2}{c}{Phase $\phi= \{c\}$}\\
\midrule
  Perfect coordination of units, no voltage unbalance constraints & 
  306.24 & 100.00 & 306.04 & 100.00 & 305.60 & 100.00 \\
  \midrule
No coordination of units, no voltage unbalance constraints & 138.87 & 45.35 & 120.10 & 39.24 & 115.56 & 37.81 \\
\midrule
No coordination of units, VUF limit of 1.0\% imposed & 138.23 & 45.14 & 101.90 & 33.30 & 110.88 & 36.28 \\
No coordination of units, VUF limit of 0.9\% imposed & 135.26 & 44.17 &	83.23 & 27.20 & 97.34 & 31.85 \\
No coordination of units, VUF limit of 0.8\% imposed & 129.73 & 42.36 & 65.81 & 21.50 & 82.83 & 27.11 \\
No coordination of units, VUF limit of 0.7\% imposed & 120.06 & 39.20 & 49.37 & 16.13 & 67.64 & 22.13 \\
No coordination of units, VUF limit of 0.6\% imposed & 103.91 & 33.93 & 32.48 & 10.61 & 53.74 & 17.59 \\
No coordination of units, VUF limit of 0.5\% imposed & 84.93 & 27.73 & 17.68 & 5.78 & 41.4 & 13.55\\
No coordination of units, VUF limit of 0.4\% imposed & 62.71 & 20.48 & 7.04 & 2.30 & 30.53 & 9.99 \\
No coordination of units, VUF limit of 0.3\% imposed & 35.37 & 11.55 & 0.83 & 0.27 & 21.06 & 6.89 \\
No coordination of units, VUF limit of 0.2\% imposed & 15.75 & 5.14 & 0.00 & 0.00 & 12.54 & 4.10 \\
No coordination of units, VUF limit of 0.1\% imposed & 3.97 & 1.30 & 0.00 & 0.00 & 3.54 & 1.16 \\
\bottomrule
\end{tabular}
\\
\label{table: load unbalance impacts 2}
\vspace{0.3em}
\footnotesize
\noindent\textit{Note:} All simulations assume an unbalanced network, with flexibility provided by single-phase units connected to different phases.
\end{table*}

The findings of this subsection emphasise the importance of considering phase unbalance impacts when assessing flexibility services in distribution networks. If these impacts are ignored, the potential of DER flexibility services may be greatly overestimated. The following subsection presents a realistic case study to demonstrate the scalability of the proposed framework.

\subsection{Case Study: 221-bus LV Distribution Network in the UK}
The proposed flexibility quantification framework is applied to a real anonymised 0.4 kV distribution network located in the North West of England \cite{churkin_Zenodo}.
The network is visualised in Fig.~\ref{Fig: 221bus scheme} as a graph using the force-directed graph layout algorithm ForceAtlas2 \cite{Jacomy2014}. The figure illustrates the network topology and the location of power demands across the network, where circles of different sizes correspond to the total apparent power consumed at each bus. The smallest circles indicate buses with no loads, while the largest circles stand for buses with a load of 3.52 kVA. The distribution of loads across phases is given in Table~\ref{table: UK network data}. The network is supplied from a single three-phase balanced source, the primary substation (displayed as a triangle in Fig.~\ref{Fig: 221bus scheme}). Yet, due to the uneven distribution of single-phase loads, the network experiences a voltage unbalance, with the highest VUF of 0.78\% over all buses. Such a low level of voltage unbalance is considered acceptable according to existing grid codes \cite{EnergyNetworksAssociation1990,NationalGrid2014}. However, this voltage unbalance can still impact the network's flexibility and limit the operation of DER.

\begin{figure*}
    \centering
    \includegraphics[width=0.98\textwidth]{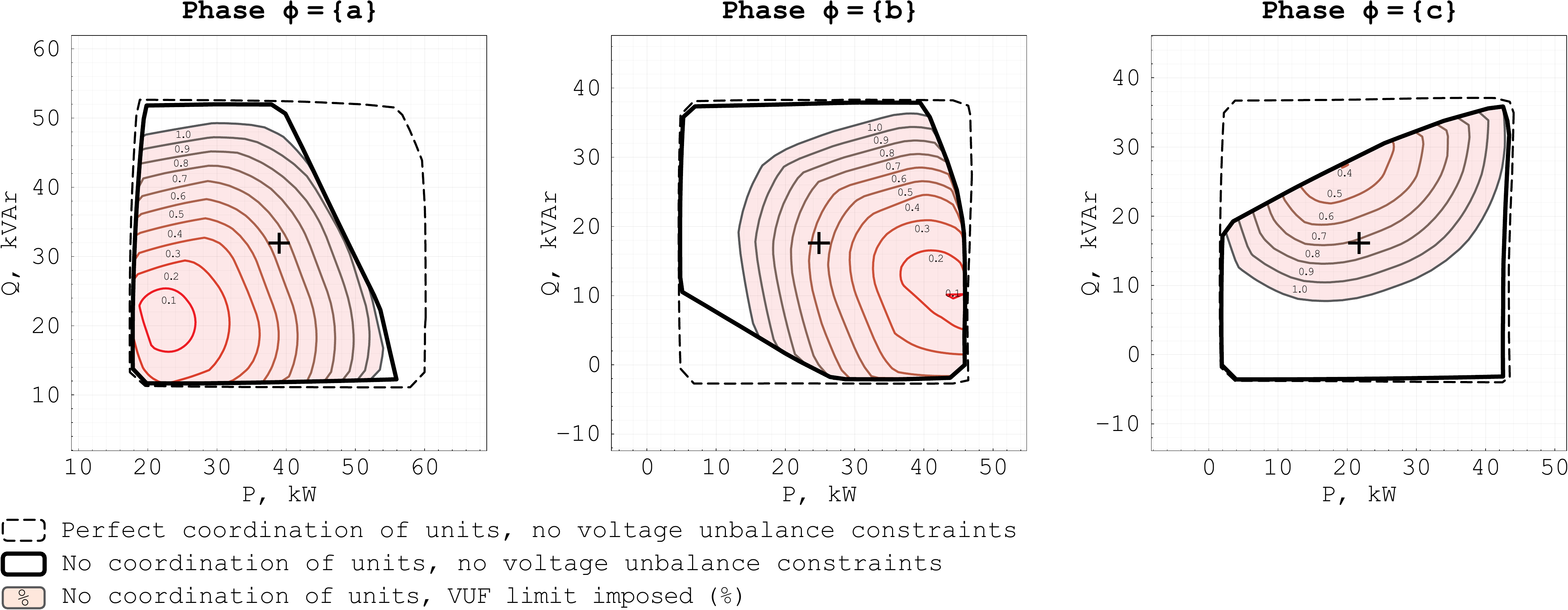}
    \caption{Impact of phase coordination and voltage unbalance constraints on flexibility services in the real 221-bus system. The aggregated P-Q flexibility areas are shown for each phase. Cross markers indicate the initial operating point with no flexibility provision, while the coordinates represent the total network power consumption per phase. Flexibility services are provided by 12 single-phase units. Imposing phase coordination constraints prevents joint flexibility provision by units connected to different phases, thereby limiting voltage unbalance management and reducing the feasible area for each phase.
    }
    \label{fig: 221bus coordination analysis}
\end{figure*}

To investigate the impact of phase unbalance on flexibility services, 12 single-phase flexible units (four units per phase) were placed in the network as shown in Fig.~\ref{Fig: 221bus scheme}. 
{\color{black}The locations of the units were selected to reflect realistic DER deployments in LV networks and to highlight the interrelation between unit operation and voltage unbalance. Specifically, the units were placed at nodes located towards the end of feeders, where voltage unbalance effects are more pronounced and flexibility injections are more likely to be constrained.}
Each unit has a P-Q capability of ±5 kW and ±5 kVAr. That is, units can independently regulate their active and reactive power output in the range of [-5.0,5.0] kW and [-5.0,5.0] kVAr.
Then, the aggregated P-Q flexibility areas were calculated per each phase by iteratively solving the flexibility estimation model \eqref{Model: objective}-\eqref{Model: phase_coordination}. The impact of phase coordination and voltage unbalance constraints on P-Q flexibility areas in the 221-bus system is shown in Fig.~\ref{fig: 221bus coordination analysis}.
{\color{black}The red shaded areas indicate simulations in which VUF constraints \eqref{Model: vuf_lim} are imposed.\footnote{\color{black}To avoid scalability issues, in the 221-bus system, VUF constraints \eqref{Model: v1}-\eqref{Model: vuf_lim} are imposed only on the 7 buses with the highest voltage unbalances and close to the flexible units, as shown in Fig.~\ref{Fig: 221bus scheme} by the black cross markers. A detailed discussion of the voltage unbalance modelling and scalability properties is provided in Section~\ref{subsection: scalability}.}
The quantitative results are summarised in Table~\ref{table: load unbalance impacts 3}.
}

Similar to the previous analysis for the simple 5-bus system, the P-Q areas in Fig.~\ref{fig: 221bus coordination analysis} indicate that the network's flexibility is maximised when all single-phase flexible units are perfectly coordinated, as shown by the dashed boundaries. However, the absence of coordination between units connected to different phases leads to significant losses in the aggregated DER flexibility potential. 
{\color{black}For example, as shown by the quantitative results in Table~\ref{table: load unbalance impacts 3}, a lack of flexible unit coordination leads to a flexibility area reduction of 15.72–29.23\%, depending on phase. When voltage unbalance constraints \eqref{Model: vuf_lim} are imposed, the reduction increases further to 34.80–60.28\%, depending on phase.
These observations suggest that at least 30\% of theoretical aggregated flexibility may not be achievable in realistic unbalanced LV distribution networks.}


\begin{table*}
\caption{\color{black}Impact of Phase Coordination and Voltage Unbalance Constraints on Aggregated P-Q Flexibility in the 221-Bus System}
\centering
\begin{tabular}{@{}l
S[table-format=4.2] @{\hspace{0.6em}(} S[table-format=3.2] @{\,\%)\hspace{1.8em}}
S[table-format=4.2] @{\hspace{0.6em}(} S[table-format=3.2] @{\,\%)\hspace{1.8em}}
S[table-format=4.2] @{\hspace{0.6em}(} S[table-format=3.2] @{\,\%)}@{}}
\toprule
 \multirow{2}{*}[-0.25em]{Simulation} & \multicolumn{6}{c}{Aggregated P-Q flexibility area per phase, kVAr\textsuperscript{2} (\%)}\\
  \cmidrule(l){2-7} 
 & \multicolumn{2}{c}{\hspace{-0.9em}Phase $\phi= \{a\}$} & \multicolumn{2}{c}{\hspace{-1.8em}Phase $\phi= \{b\}$} & \multicolumn{2}{c}{Phase $\phi= \{c\}$}\\
\midrule
  Perfect coordination of units, no voltage unbalance constraints & 
  1721.72 & 100.00 & 1709.93 & 100.00 & 1702.88 & 100.00 \\
  \midrule
No coordination of units, no voltage unbalance constraints & 1211.63 & 70.37 & 1441.21 & 84.28 & 1300.69 & 76.38 \\
\midrule
No coordination of units, VUF limit of 1.0\% imposed & 1122.60 & 65.20 & 1050.17 & 61.42 & 676.34 & 39.72 \\
No coordination of units, VUF limit of 0.9\% imposed & 1035.54 & 60.15 & 935.33 & 54.70 & 517.11 & 30.37 \\
No coordination of units, VUF limit of 0.8\% imposed & 915.38 & 53.17 & 822.88 & 48.12 & 376.33 & 22.10 \\
No coordination of units, VUF limit of 0.7\% imposed & 787.25 & 45.72 & 710.63 & 41.56 & 253.27 & 14.87 \\
No coordination of units, VUF limit of 0.6\% imposed & 662.87 & 38.50 & 594.61 & 34.77 & 146.07 & 8.58 \\
No coordination of units, VUF limit of 0.5\% imposed & 544.67 & 31.63 & 476.56 & 27.87 & 58.20 & 3.42\\
No coordination of units, VUF limit of 0.4\% imposed & 434.67 & 25.25 & 357.13 & 20.89 & 0.02 & 0.00 \\
No coordination of units, VUF limit of 0.3\% imposed & 330.72 & 19.21 & 211.44 & 12.37 & 0.00 & 0.00 \\
No coordination of units, VUF limit of 0.2\% imposed & 198.74 & 11.54 & 72.17 & 4.22 & 0.00 & 0.00 \\
No coordination of units, VUF limit of 0.1\% imposed & 56.18 & 3.26 & 0.69 & 0.04 & 0.00 & 0.00 \\
\bottomrule
\end{tabular}
\\
\label{table: load unbalance impacts 3}
\end{table*}


Note that in this simulation setting, all flexible units have been assigned the same P-Q capabilities and distributed evenly across phases. Yet, their operation is limited considerably due to phase coordination issues and voltage unbalance. It follows that the provision of flexibility services by DER can be challenging even in LV networks with low load unbalance as flexible units can exacerbate existing voltage unbalances by producing or consuming power. To effectively manage such unbalances and voltage constraints, units connected to different phases must be perfectly coordinated. The simulations performed for the realistic 221-bus network demonstrate that lack of coordination between units significantly reduces the potential of DER flexibility services and can magnify voltage unbalance problems. Other aspects and implications of these findings are discussed in Section \ref{Section: discussion}.

\subsection{Stochastic Analysis}
Previous simulations in this section focused on a deterministic analysis of aggregated P-Q flexibility in LV unbalanced networks. The deterministic setting enabled an interpretable analysis of the impacts of phase unbalance and DER coordination on flexibility provision, with clear visualisations. This subsection complements the reported findings by providing a stochastic analysis of voltage unbalances caused by unbalanced loads and flexible units. Specifically, the aim is to estimate how often VUF limit violations can occur when uncoordinated flexible resources operate in unbalanced networks. For this purpose, 1,000 load scenarios were randomly generated for the 221-bus LV distribution network. In each scenario, P-Q power demand values were varied independently across all buses by ±50\% (sampled from a uniform distribution) to reflect temporal and spatial variability.

Then, to simulate the operation of flexible units, the following flexibility configurations were considered:
\vspace{2cm}

\begin{itemize}
\item \textit{No flexibility provision}: all units are switched off. This baseline solution corresponds to the intrinsic voltage imbalances in the network due to unbalanced loads.
\item \textit{Uncoordinated flexibility provision}: all units operate independently and without coordination. Each unit can consume or generate power. The values of flexible power injections are generated randomly (uniformly) for each scenario within the capacity limits of the units.
\item \textit{Power consumption by phase A}: only units connected to phase A are active, all consuming power. This stresses the system and exacerbates existing voltage unbalances in the network. The values of flexible loads are generated randomly (uniformly) for each scenario within the capacity limits of the units.
\end{itemize}

\begin{figure}
    \centering
    \includegraphics[width=0.9\columnwidth]{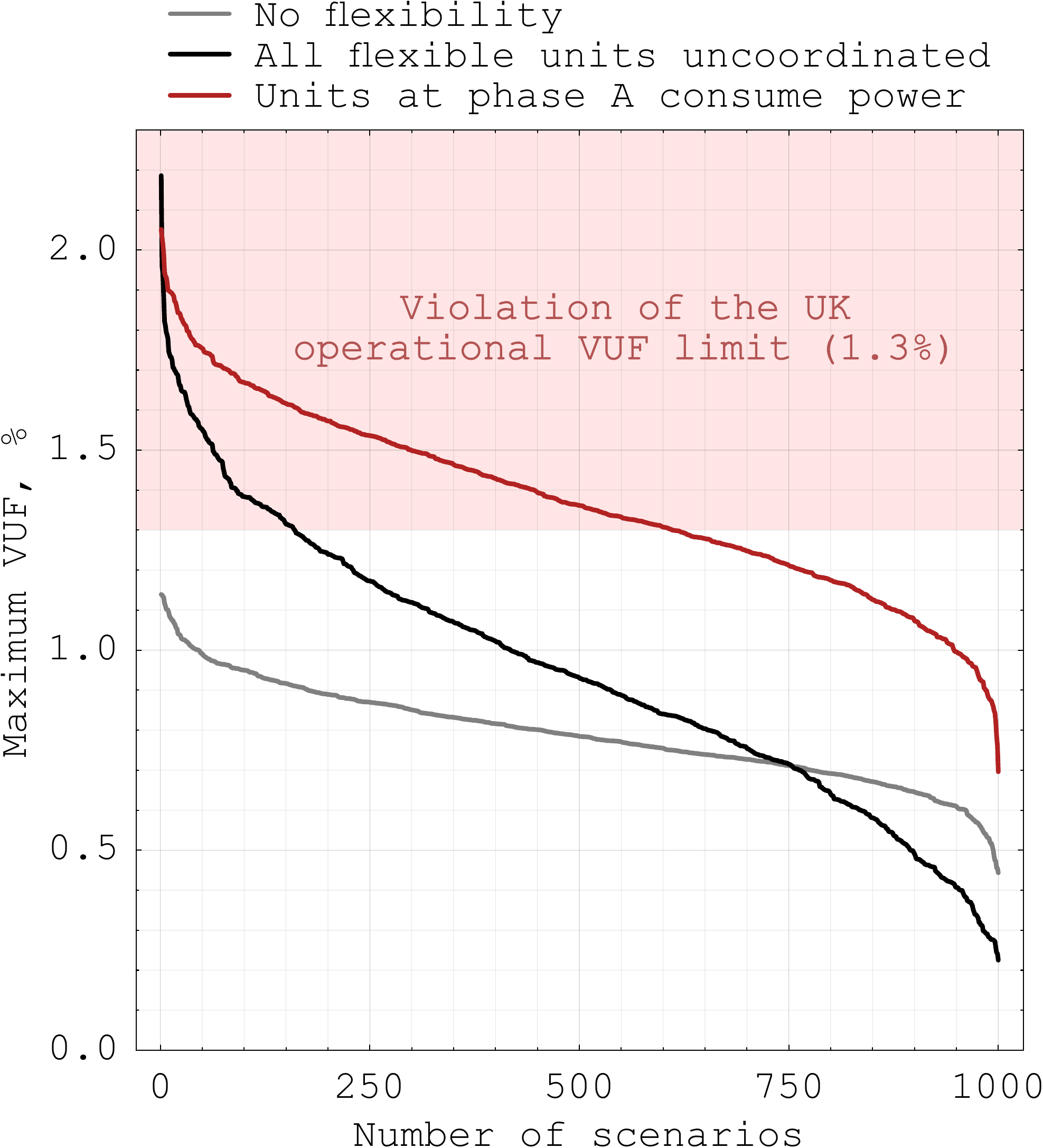}
    \caption{Sorted maximum VUF values (VUF duration curves) for 1,000 stochastic load scenarios under three flexibility configurations: (i) no flexibility provision, (ii) uncoordinated flexible power consumption/generation by all units, and (iii) power consumption only by units connected to phase A. The shaded red area highlights violations of the UK operational VUF limit of 1.3\% for LV networks.}
    \label{Fig: VUF stochastic analysis}
\end{figure}

For each scenario, the OPF problem \eqref{Model: objective}-\eqref{Model: qf_min_max} was solved, and the maximum voltage unbalance across all buses was calculated. The results are presented in Fig.~\ref{Fig: VUF stochastic analysis}, where the VUF values are sorted in descending order, thus forming VUF duration curves across the scenarios. Note that without flexibility, unbalanced loads do not cause significant voltage unbalances – maximum VUF values are well below 1.3\% for all scenarios. However, when introducing uncoordinated flexible resources, a significant share of scenarios (over 15\%) exhibit violations of the voltage unbalance limit. These results highlight the importance of phase-aware control strategies for flexible resources: without coordination, flexibility provision can inadvertently introduce harmful voltage unbalances.

\section{Discussion}\label{Section: discussion}
The simulations performed in Section \ref{Section: results} demonstrate the impacts of phase unbalance and imperfect coordination between distributed flexibility providers. Specifically, it is shown that the total DER flexibility potential can be significantly reduced due to voltage unbalance constraints and lack of coordination between flexible units connected to different phases.
This highlights the importance of accurate nonlinear three-phase OPF models for the analysis of the network's aggregated P-Q flexibility, as such models are capable of simulating voltage unbalances and coordination issues across phases.
In this regard, the proposed flexibility quantification framework allows to directly translate voltage unbalance and phase coordination constraints as reductions of the aggregated P-Q flexibility. Thus, it can be used by DSO to realistically predict flexibility available in distribution networks and avoid severe voltage unbalances.

{\color{black}Yet, the proposed framework relies on several assumptions regarding the modelling of DER operation in unbalanced distribution networks. {\color{black}This section further discusses the implications of these assumptions, including the choice of DER placement locations, voltage unbalance definitions, DER coordination limitations and communication delays, as well as multi-period flexibility aggregation. 
The section concludes with a series of scalability tests, elaborating on the computational complexity of the framework.}

\subsection{Impact of DER Location on Voltage Unbalance and Losses}
Multiple aspects of flexibility provision in unbalanced LV distribution networks have to be further investigated. One of the key factors affecting the aggregated P-Q flexibility is the relationship between the placement of flexible units and locations of voltage unbalances. For instance, in the 221-bus case study displayed in Fig.~\ref{Fig: 221bus scheme}, the highest load unbalance is happening at the feeder with the flexible unit \texttt{F5\{a\}} connected. Six more flexible units are placed in this part of the network. Therefore, the design of this experiment inherently creates constraints on the operation of these flexible units due to the nearby voltage unbalances. That is, units must be coordinated when consuming or producing power to manage voltage constraints and keep the network's operation feasible. For example, units connected to phase $\phi= \{a\}$ (the phase with the highest load) must not increase their power consumption while units in other phases consume power. Otherwise, voltage unbalance in some parts of the network may increase to unacceptable levels or even the voltage limits \eqref{Model: V_min_max} for phase $\phi= \{a\}$ may be violated.
Note that this would not happen if flexible units were located far from the load unbalance, say, at the primary substation. Therefore, the question arises: how does the initial network unbalance impact the network's P-Q flexibility and how does it restrict flexible resources at different locations?
This question has implications for DER planning models. Depending on DSO metrics and objectives, some DER should be placed closer to the primary substation, while other flexible resources should be located closer to existing voltage unbalances to manage voltage constraints.
A similar problem was considered in \cite{Wang2021,Antic2024}, where it was demonstrated that suboptimal selection of the DER connection phase can lead to voltage unbalance and limit DER capacity.

{\color{black}
The provision of flexibility services can also be impacted considerably by the integration of EV charging stations. This impact can be especially pronounced in residential and workplace areas, where charging points are predominantly single-phase. In such scenarios, uncoordinated EV charging may reduce available flexibility due to tighter voltage and VUF constraints. Investigating EV-driven unbalance and its implications for flexibility aggregation is a valuable direction for future research.
}

Another important aspect of flexible resource placement is its impact on power losses. In the 221-bus system, the total active power losses amount to 2.4 kW (2.8\%) for the initial operating point with no flexibility provision. The operation of flexible units can either reduce or exacerbate these losses, depending on their location and coordination. 
{\color{black}
Note that the loss-minimisation problem is distinct from the aggregated P-Q flexibility estimation. The loss-minimisation objective does not aim to find the limits of the P-Q area but instead optimises the system operation at a given feasible point within the area. From a modelling perspective, this requires fixing the system's operating point at the reference bus, $s_{i=i_{\text{ref}},\phi} = s_{i=i_{\text{ref}},\phi}^\prime$, and defining the dispatch optimisation objective function, such as minimisation of total active power losses or the total cost of flexibility provision. Similar cost-optimisation problems within the P-Q area have been studied in \cite{Sarstedt2022, Churkin2024}. Due to network constraints and heterogeneous unit costs, the optimal dispatch solutions and the associated losses can differ significantly even between nearby operating points in the P-Q space.
}

\subsection{Unbalance Definition and Modelling Assumptions}
{\color{black}In this work, voltage unbalance in LV distribution networks is quantified and constrained using the VUF, defined as the ratio between the negative sequence and positive sequence of voltage components. This definition is widely adopted for electric power quality monitoring and is recommended by standards such as IEC 61000-2-2 and IEEE 1159 \cite{ nakadomari2024}. However, alternative definitions of voltage unbalance have been proposed in the literature. For example, voltage unbalance can be formulated as the ratio of the zero sequence to positive sequence. Voltage unbalance can also be quantified through the phase voltage unbalance rate (PVUR), computed based on the average phase voltage magnitude and the maximum deviation among phases. In addition, the current unbalance can be analysed using phase current magnitudes and current sequence components. It is therefore important for future research to systematically compare alternative unbalance definitions and analyse their impact on P-Q flexibility aggregation.

Another key modelling assumption made in this work is the use of a three-phase three-wire AC OPF formulation. This formulation captures the propagation of voltage unbalance along network lines and allows the explicit modelling of single-phase, two-phase, and three-phase load connections. However, it does not explicitly model the neutral conductor and grounding effects. Therefore, the proposed framework can be extended to four-wire network models. Analysis of the impact of the neutral conductor and grounding on aggregated P-Q flexibility is left for future research.

\subsection{DER Coordination and Communication Constraints}
{\color{black}A few modelling assumptions are made in this work regarding data availability, network observability, and DER coordination. Specifically, to demonstrate the fundamental impact of voltage unbalance constraints on flexibility aggregation, two operating conditions are compared: 1) a perfectly coordinated case, in which all DER are jointly optimised across phases, and 2) an uncoordinated case, in which DER connected to different phases are optimised independently, as imposed by constraint \eqref{Model: phase_coordination}. Both conditions are idealised. The “perfect coordination” assumption is not always valid in real LV distribution networks, as some units may not be observable or may experience communication delays. Conversely, the “no coordination” assumption is overly restrictive as it implies that units in other phases are inactive. 

{\color{black}
In practice, only a subset of flexible units may be coordinated, while other units may experience communication delays or be unavailable for certain time periods.
The proposed P-Q flexibility assessment framework can represent such situations by introducing a set of controllable flexible units, $\mathcal{F}_c \subseteq \mathcal{F}$, whose P-Q capabilities are available for flexibility aggregation. That is, whenever a unit leaves the flexibility market or experiences observability or communication problems, it is excluded from $\mathcal{F}_c$, which reduces the aggregated flexibility available for that period.
Some units may also inject or withdraw power due to their own purposes without providing flexibility services for the system. In such cases, power setpoints of the units become inputs to the framework rather than control variables.

Note that communication delays are inherently a temporal phenomenon: a delayed unit responds later than units with lower communication latency. Therefore, to accurately assess the impact of delays on aggregated P-Q flexibility, dynamic multi-period modelling is required. Such modelling can explicitly track units’ setpoints across consecutive time periods and capture how delayed control actions affect feasibility over time. The implications of multi-period flexibility modelling are discussed further in the next subsection.
}

{\color{black}
\subsection{Multi-Period Flexibility Aggregation}\label{subsection: multiperiod}
The proposed steady-state multi-phase P-Q flexibility estimation model \eqref{Model: objective}-\eqref{Model: phase_coordination} characterises the limits of aggregated P-Q flexibility at a single time snapshot. 
This single-period formulation is intentionally selected to isolate and clearly demonstrate the impact of voltage unbalance and phase coordination constraints on aggregated flexibility. However, this model does not account for intertemporal constraints or the dynamics of flexibility provision.

The need for multi-period flexibility modelling is acknowledged in the literature \cite{Avramidis2021,Petrou2021,Liu_Ochoa2022,Capitanescu_barriers,Muller2019}. Such models capture time-varying operating limits for DER, e.g., through dynamic operating envelopes defined on an hourly basis.
The need for time-varying limits is also recognised in studies on aggregated P-Q flexibility \cite{Riaz2022,Contreras_time_based,Lopez_multiperiod}. A common approach is to recompute the P-Q flexibility area for a sequence of time periods, tracking the area shifts due to changing loads, network conditions, and DER setpoints.
Importantly, each area cannot be treated independently, as the operation of energy-constrained resources such as BESS and EV couples decisions and constraints across different periods.

Specifically, compared to the steady-state formulation, multi-period modelling introduces the following challenges:
\begin{itemize}
\item \textit{Size of the problem}: all variables and constraints of the single-period formulation \eqref{Model: objective}-\eqref{Model: phase_coordination} are replicated for every period $t \in \mathcal{T}$, so the problem size grows proportionally to the length of the operating horizon. For example, control variables of each flexible unit become time-dependent, $P_{f,i,\phi,t}$ and $Q_{f,i,\phi,t}$, denoting active and reactive power output at period $t$. 
\item \textit{Intertemporal coupling}: the operation of energy-constrained units is linked across periods through their state of charge. Each unit requires an additional energy variable, $E_{f,i,\phi,t}$, that evolves with its charging and discharging actions and the length of the periods, e.g., $E_{f,i,\phi,t}=E_{f,i,\phi,t-1}-\Delta t P_{f,i,\phi,t}$. This variable is bounded by the unit's energy capacity, $E_{f,i,\phi}^{\min} \leq E_{f,i,\phi,t} \leq E_{f,i,\phi}^{\max}$. Moreover, the successive changes in unit's power output are constrained by its ramp-up and ramp-down limits, e.g., $-R_{f,i,\phi}^{\downarrow} \leq P_{f,i,\phi,t}-P_{f,i,\phi,t-1} \leq R_{f,i,\phi}^{\uparrow}$.
\item\textit{Path dependence}: as a result of this coupling, the aggregated P-Q flexibility area at a given period is no longer unique. It depends on the initial state of charge of units, the length of the operating horizon, and unit decisions in the preceding and following periods \cite{Lopez_multiperiod}. 
\end{itemize}

Considering these aspects, the single-period P-Q flexibility area can be interpreted as an optimistic case, in which no energy is reserved for adjacent periods. Thus, the steady-state estimation provides an upper bound on the flexibility achievable under temporal coupling. In realistic multi-period applications, time-coupling represents another practical barrier to flexibility aggregation, as units must be coordinated not only across phases but also across periods of the operating horizon.

Note that the proposed framework does not aim to optimise the operation of flexible resources across multiple periods. Instead, the setpoints of units, together with their state of charge and ramp limits, produce a time-varying input for the model. Based on this input, the framework estimates the aggregated P-Q flexibility and detects voltage unbalance violations for each period. A dedicated multi-period formulation with optimal control of flexible units over the operating horizon is left for future research.
}

\subsection{Computational Complexity and Scalability}\label{subsection: scalability}
{\color{black}
The main computational burden of multi-phase P-Q flexibility estimation comes from the need to iteratively solve the nonlinear optimisation problem \eqref{Model: objective}-\eqref{Model: phase_coordination}. As illustrated in Fig.~\ref{Fig: iterative PQ algorithm}, the number of optimisation problems to be solved depends on the desired granularity of P-Q area estimation, i.e., the number of feasible operating points found at the boundary of the area. Thus, the scalability of the proposed framework depends on the computation time of the underlying AC OPF problems, which are driven by the following factors:
\begin{itemize}
\item Distribution network size (number of buses and lines).
\item Number of flexible units available in the network.
\item Severity of voltage unbalance constraints \eqref{Model: vuf_lim} and the number of VUF-constrained buses $\mathcal{N}^{\text{vu}}$.
\item Relationship between the placement of flexible units, existing voltage unbalances, and VUF-constrained buses.
\end{itemize}

This subsection elaborates on these factors through a series of scalability tests, offering further insight into the computational complexity and scalability of the framework. All tests were performed on a laptop with an Intel Core i7-1365U CPU 1.80 GHz and 16 GB of RAM.
Detailed configurations of the tests, together with the simulation code, timing data, and visualisations, are available in the repository \cite{3FlexAnalyser}.
The results suggest that the framework can scale to larger and more complex networks. Importantly, because the underlying AC OPF is a non-convex nonlinear problem, IPOPT solve times and convergence are governed not only by problem size but also by solver iteration counts, local nonlinearity, and the set of binding constraints.

As expected, the size of the network is a primary driver of computation time: the mean convergence time for a single OPF problem with voltage unbalance constraints is approximately 0.1 s and 40 s for the 5-bus and 221-bus cases, respectively. This suggests that in the 221-bus system, the boundary of an aggregated P-Q flexibility area defined by 50 points can be estimated in 2000 s (33 min) on average. The flexibility estimation process can be greatly sped up by using more powerful hardware and parallel computing, as independent OPF problems can be distributed across different cores and machines. Therefore, DSO can perform this flexibility assessment close to operational timescales, e.g., updating flexibility areas during intraday markets every 15 minutes. Yet, it is important to highlight that the proposed framework is not intended for real-time control. Instead, it is best suited for operational planning and market-related applications.

The next major computational factor is the number of flexible units in the network. To investigate this, scalability tests were performed on the 221-bus system for increasing numbers of units. Single-phase units are added one by one following the order specified in \cite{3FlexAnalyser, churkin_Zenodo}: units 1–4 connected to phase $\phi= \{a\}$, units 5–8 to phase $\phi= \{b\}$, and units 9-12 to phase $\phi= \{c\}$. For each test, the three-phase flexibility estimation model \eqref{Model: objective}-\eqref{Model: phase_coordination} was solved 100 times, corresponding to 100 boundary points of the P-Q area in phase $\phi= \{a\}$. A VUF limit of 1.0\% was imposed on the 7 preselected buses across all tests, and the phase coordination constraints \eqref{Model: phase_coordination} were inactive. The resulting OPF computation times are visualised in Fig.~\ref{Fig: scalability num units}, showing the distribution and median per test.
The median computation time grows substantially with the number of units, increasing by roughly an order of magnitude between 1 and 10 units. However, the growth is non-monotonic and exhibits high variance, reflecting the sensitivity of interior-point iteration counts to constraint activity and feasibility in this non-convex AC OPF problem. Importantly, the computation cost does not grow exponentially or combinatorially with the number of units. The cost even starts declining when adding units 11 and 12. This indicates that the computational complexity is driven not only by the number of units but also by their phase connection and placement. For example, the units added last in this test are connected to phase $\phi= \{c\}$. These units balance the flexibility across phases, improving the conditioning of the optimisation problem and reducing solver iteration counts.

To further investigate this scalability behaviour, a modified 221-bus system with a larger number of units was tested. Specifically, 50 single-phase units were placed in the network, allocated almost equally across the three phases. The first 12 units were connected to the same buses as in the original case study. All units were assumed to have a P-Q capability of ±1~kW and ±1~kVAr, so that the total capability is comparable to the original case. Then, a 100-point P-Q boundary was estimated. The computation times for this test lie in the range of 10-119 s, with a median of 20 s. Despite the fivefold increase in the number of units, the optimisation model remained tractable and exhibited no convergence failures. Therefore, this test demonstrates that the framework can be applied reliably to cases with larger numbers of units.
\vspace{1cm}
}

{\color{black}
Another important scalability factor is the severity of the imposed VUF limits. To investigate this, a scalability test was carried out on the 221-bus system for VUF limits varying between 1.5\% and 0.5\%.
As in the previous test, the three-phase flexibility estimation model \eqref{Model: objective}-\eqref{Model: phase_coordination} was solved 100 times for each limit, with VUF constraints imposed on the 7 preselected buses and phase coordination constraints \eqref{Model: phase_coordination} kept inactive.
Note that the problem dimension is identical across all tests, and only the tightness of the VUF constraints changes.
The resulting computation times are shown in Fig.~\ref{Fig: scalability vuf severity}.
No clear trend can be observed in these simulations: tightening the VUF limit does not produce a consistent increase in computation time. The time shifts between the tests are non-monotonic and exhibit high variance, which again highlights the interior-point solver's sensitivity to binding constraints and their relationship to unit placement. The maximum observed time increase is roughly twofold across the range of VUF limits.
This indicates that enforcing stricter VUF limits at a given set of buses does not incur a significant computational penalty, with solve times remaining within the same order of magnitude.

Finally, the set of VUF-constrained buses and its relationship to the placement of flexible units is crucial for both the framework's scalability and the solver's performance. In the original 221-bus case study, VUF constraints were imposed only on 7 preselected buses: locations with high voltage unbalances, close to the flexible units.
Constraining this small set of buses is computationally cheap and sufficient to capture the binding unbalances. Imposing VUF constraints on larger sets of locations, e.g., all 221 buses, deteriorates the performance of the interior-point solver and may lead to infeasibility. 
For example, due to large numbers of VUF constraints, IPOPT may terminate at a point of local infeasibility while trying to reach the P-Q area boundary.
Importantly, this behaviour is driven not only by the number of constraints but also by their locations in the network.
This motivates future research to develop more efficient and robust P-Q flexibility estimation models for LV unbalanced networks. Specifically, tuning IPOPT parameters should be investigated for cases with multiple VUF-constrained locations. Alternative solvers can also be tested and evaluated against IPOPT. Furthermore, the flexibility estimation model can be improved via alternative OPF formulations, approximations, and relaxations.

}

\begin{figure}
    \centering
    \includegraphics[width=0.95\columnwidth]{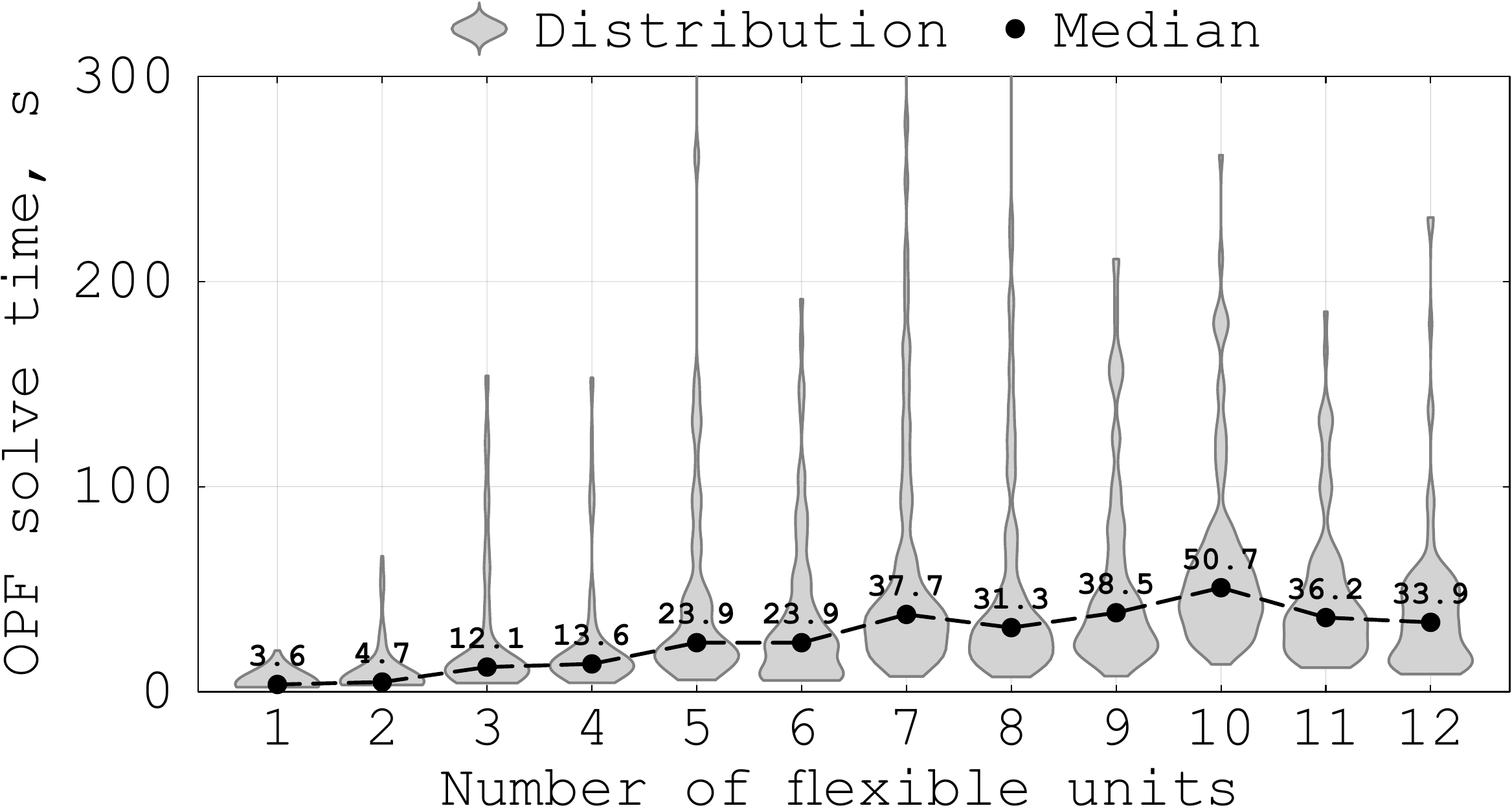}
    \caption{\color{black}Scalability of the three-phase AC OPF model for P-Q flexibility aggregation with respect to the number of flexible units in the 221-bus system.}
    \label{Fig: scalability num units}
\end{figure}

\begin{figure}
    \centering
    \includegraphics[width=0.95\columnwidth]{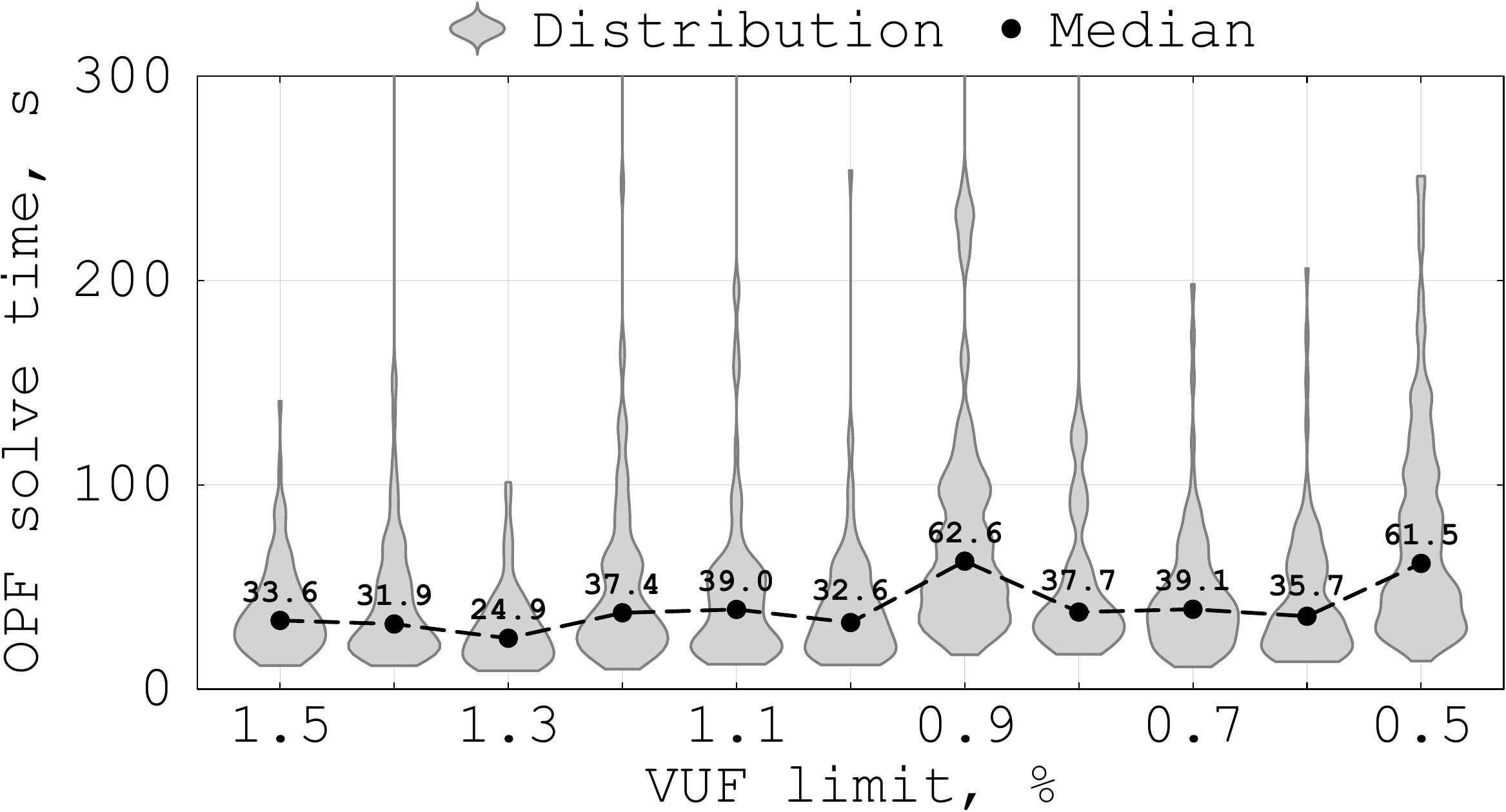}
    \caption{\color{black}Scalability of the three-phase AC OPF model for P-Q flexibility aggregation with respect to the VUF limit value in the 221-bus system.}
    \label{Fig: scalability vuf severity}
\end{figure}

\section{Conclusion}\label{Section: conclusion}
This paper investigates the provision of flexibility services in unbalanced LV distribution networks and proposes a framework for quantifying the impacts of phase unbalance and imperfect coordination between distributed flexibility providers.
At the core of the framework lies an accurate nonlinear three-phase AC OPF model designed for simulating the operation of flexible units connected to different phases.
To characterise feasible flexibility services available in networks, the concept of aggregated P-Q flexibility areas is applied.
The combination of these models and concepts allows to directly translate voltage unbalance and phase coordination constraints into reductions in aggregated P-Q flexibility.
The simulations, performed for an illustrative 5-bus distribution system and a real 221-bus LV network in the UK, demonstrate that a significant share (over 30\%) of total aggregated DER flexibility potential may be unavailable due to voltage unbalances and lack of coordination between DER connected to different phases.

A detailed analysis of these simulations enables to formulate several findings explaining the operation of DER in unbalanced distribution networks and their role in aggregated flexibility services.
For example, it is found that flexibility services in LV distribution networks can be significantly constrained due to inherent load unbalances, even in cases where flexible resources provide balanced output.
Moreover, in cases where flexible units are connected to different phases, they can exacerbate existing voltage unbalances by producing or consuming power. To effectively manage such unbalances and voltage constraints, units connected to different phases must be perfectly coordinated.
However, in realistic networks, flexible units cannot effectively coordinate to manage voltage constraints across different locations and phases. Thus, provision of some flexibility services by the network may be infeasible or may lead to severe voltage unbalances.

Future research can further explore different aspects of flexibility aggregation in unbalanced LV distribution networks. For example, as the location of flexible units plays a crucial role in managing voltage unbalances, it is worth exploring flexibility planning models and metrics for optimising DER placement, capturing the relationship between DER locations and buses with the highest load unbalances.
{\color{black}It is also important to investigate EV-driven unbalance arising from single-phase charging stations and its implications for flexibility aggregation.
}
From a computational perspective, the framework can be further improved to better scale with the number of flexibility providers and VUF constrained buses.
{\color{black}Finally, as this work demonstrates the technical necessity of cross-phase coordination, an important future research direction is the development of market designs and incentive mechanisms that could enable such coordination economically in real distribution systems.}

\ifCLASSOPTIONcaptionsoff
  \newpage
\fi



%



\bibliographystyle{IEEEtran}
\bibliography{references.bib}

%








\end{document}